\documentclass[3p]{elsarticle}

\usepackage[T1]{fontenc}
\usepackage[utf8]{inputenc}

\usepackage[T1]{fontenc}
\usepackage[utf8]{inputenc}
\usepackage{hyperref}
\usepackage{amsmath}
\usepackage{amssymb}
\usepackage{xcolor}

\graphicspath{.}

\newcommand{\tN}[2]{\mathcal{\widehat{N}}\left(#1, #2\right)}

\journal{Journal of Medical Internet Research}

\begin{document}

	\begin{frontmatter}
		\title{Social Contagion in COVID-19 Discussions within the Belgian Reddit Community: A Statistical and Modeling Study}

		\author[1]{Tim Van Wesemael\corref{cor1}}
		\ead{tim.vanwesemael@ugent.be}
		\author[2,3]{Luis E.C. Rocha}
		\ead{luis.rocha@ugent.be}
		\author[1,4,5]{Tijs W. Alleman}
		\ead{twa27@cornell.edu}
		\author[1]{Jan M. Baetens}
		\ead{jan.baetens@ugent.be}

		\cortext[cor1]{Corresponding author}

		\affiliation[1]{%
			organization={BionamiX, Department of Data Analysis and Mathematical Modelling, Ghent~University},
			addressline={Coupure~Links~653},
			postcode={9000},
			city={Ghent},
			country={Belgium},
		}
		\affiliation[2]{%
			organization={Department of Economics, Ghent~University},
			addressline={Sint-Pietersplein~5},
			postcode={9000},
			city={Ghent},
			country={Belgium},
		}
		\affiliation[3]{%
			organization={Department of Physics and Astronomy, Ghent~University},
			addressline={Proeftuinstraat~86},
			postcode={9000},
			city={Ghent},
			country={Belgium},
		}
		\affiliation[4]{%
			organization={Department of Public and Ecosystem Health, College of Veterinary Medicine, Cornell University},
			addressline={Ithaca},
			postcode={New York~14853-6401},
			country={United States},
		}
		\affiliation[5]{%
			organization={Department of International Health, Johns Hopkins Bloomberg School of Public Health},
			addressline={Baltimore},
			postcode={Maryland~21205},
			country={United States},
		}


		\begin{abstract}

			Understanding how sentiment toward COVID-19 mitigation measures evolves on social media can inform both epidemiological models and public health policy. We analyzed 655,642 posts by 28,559 users on r/Belgium from January 2020 to June 2022, classifying posts into three mitigation topics (\emph{lockdowns}, \emph{masks}, \emph{vaccinations}) using a BERT-based topic model and scoring sentiment with a RoBERTa-based classifier.
			Post volume tracked external events such as policy announcements, but we found no evidence of within-Reddit social contagion in topic initiation, suggesting topics are seeded by external information rather than platform-internal spread.
			Sentiment, however, exhibited significant homophily: comment sentiment correlated with that of the parent post.
			To capture the underlying dynamics, we developed the Smooth Latent-Expressed Bounded Confidence (SLEBC) model, which distinguishes a latent sentiment trajectory from noisy expressed sentiment and uses bounded confidence rather than linear update rules.
			Evaluated against two alternatives by WAIC, SLEBC fit best across all three topics.
			The model indicates that expressed sentiment adapts more strongly to the immediate parent comment than the user's latent state updates from interaction history, suggesting that expressed sentiment is a poor proxy for underlying opinion.
			These findings imply that infodemic models for Reddit-like platforms should seed topics from external sources and model sentiment spread via bounded confidence mechanisms.
			\\\\
			\textbf{Structured Abstract}

			\textbf{Background:}
			Understanding how sentiment toward COVID-19 mitigation measures evolves on social networks can inform infectious disease models and policymakers.
			Despite numerous studies describing social media interactions during the pandemic, few have modeled the underlying dynamics of sentiment contagion and polarization.

			\textbf{Objective:}
			We investigated topic emergence and sentiment evolution in COVID-19 mitigation discussions on r/Belgium, focusing on (i) whether discussion topics exhibited social contagion, (ii) whether expressed sentiment displays homophily, and (iii) how this homophily forms under local interactions and long-term trends.

			\textbf{Methods:}
			We classified posts on r/Belgium between 1 January 2020 and 30 June 2022 into three mitigation topics (\emph{lockdowns}, \emph{masks}, \emph{vaccinations}) using a pretrained BERT topic model, and assigned English posts a sentiment using a RoBERTa-based classifier.
			We examined temporal patterns of post volume and tested for social contagion in topic initiation.
			Sentiment homophily was quantified by comparing observed comment-parent sentiment pairs to null distributions.
			The novel Smooth Latent-Expressed Bounded Confidence (SLEBC) model dynamically captured sentiment evolution, distinguishing between latent sentiment trajectories and noisy expressed sentiment.
			We tested the model against two alternatives, one with a linear update and one without latent state, using the Watanabe-Akaike Information Criterion (WAIC).

			\textbf{Results:}
			Analysis of 655,642 posts by 28,559 users revealed that post volume was associated with external events such as policy announcements and media reports.
			There was no evidence of within-Reddit social contagion in topic initiation.
			However, sentiment exhibited significant homophily, with comment sentiment correlating with parent sentiment.
			The SLEBC model reproduced observed sentiment patterns (WAIC: -28.5 to -18.4 across topics), outperforming both alternatives (-21.1 to -17.4 for the linear model and 6.8 to 692 for the one without latent state).
			It slightly underestimated sentiment homophily, but still outperformed the alternatives in this regard.
			In the SLEBC model, expressed sentiment adapted more strongly to the immediate parent comment than the user's latent state updated based on interaction history (proportions of users showing this pattern: 0.74, 0.70, and 0.51 for \emph{lockdowns}, \emph{masks}, and \emph{vaccination}).

			\textbf{Conclusions:}
			Discussion topics on r/Belgium are not driven by within-platform social contagion, but sentiment dynamics are shaped by within-thread interactions.
			The SLEBC model suggests users adapt expressed sentiment to match the post they reply to, highlighting that expressed sentiment may poorly reflect underlying latent sentiment.
			Infodemic models for Reddit-like platforms could benefit from incorporating external information sources for topic seeding and from using bounded confidence rather than linear contagion mechanisms for sentiment spread.
		\end{abstract}
	\end{frontmatter}

	\section{Introduction}\label{sec:introduction}

	To mitigate the burden on healthcare systems during a pandemic, governments may be forced to introduce mandatory measures.
	The COVID-19 pandemic served as a prime example, whose myriad of mitigation measures had far reaching societal consequences~\cite{Piette2022, Reiriz2023, Alleman2026}.
	Initially in 2020, these were mainly non-pharmaceutical interventions, such as lockdowns, contact reductions, and mask mandates~\cite{BroekAltenburg2021, Luyten2022}.
	As the pandemic progressed, vaccination campaigns were initiated near the end of 2020, and the focus gradually shifted to vaccine distribution during 2021.
	The success of these interventions depended not only on their epidemiological effectiveness but also on public adherence, which was influenced by individual attitudes, information exposure, and social interactions.
	Understanding public sentiment toward these measures is therefore crucial for health informatics, as it can inform epidemiological models, guide public health communication strategies, and support evidence-based policymaking~\cite{West2020, Krawczyk2021}.

	An unprecedented surge of health-related information on social media accompanied the pandemic, leading to what has been termed an ``infodemic'', the rapid spread of both accurate and inaccurate information that can hinder effective public health responses~\cite{Cinelli2020}.
	This phenomenon has given rise to infodemiology, the science of monitoring and analyzing digital health information to inform public health practice~\cite{Cinelli2025}.
	Social media platforms have become essential data sources for infodemiology, enabling real-time monitoring of public health sentiment and tracking information diffusion patterns~\cite{Alamoodi2021}.
	Research has demonstrated that social media discussions about health topics are influenced by multiple factors.
	While traditional media coverage and official announcements drive much of the online discourse~\cite{Kurten2021, Park2024}, the spread of information within social networks also plays a crucial role.
	Studies that have established correlations between sentiment expressed online and vaccine uptake, highlight the potential of social media data as a proxy for public health attitudes~\cite{Lyu2022, Cheng2023}.

	However, infodemics are complex and multifaceted, as sentiment expressed on social media can reflect both support and opposition, and discussions are shaped by misinformation spread~\cite{Pierri2023}, echo chambers~\cite{Cinelli2021}, and platform-specific dynamics.
	Twitter (now X) has been the most extensively studied platform for COVID-19 sentiment analysis~\cite{Alamoodi2021, Medford2020, Lanier2022, Xie2023b}, including research specific to Belgium on monitoring emotions~\cite{Kurten2021} and support for specific mitigation measures~\cite{Scott2021}.

	Reddit, with its structured discussion threads and pseudo-anonymous communities, provides a different informative environment for studying online health discussions~\cite{Medvedev2019, Corradini2024}.
	Unlike Twitter's brief posts, Reddit enables extended, threaded conversations.
	Previous research on COVID-19-related subreddits has documented both temporal persistence~\cite{Melton2021} and shifts~\cite{Liu2021} in sentiment, regional differences in public concerns~\cite{Yurtsever2023, Yan2021}, and the role of external events in driving discussion volume and emotions~\cite{Park2024, Basile2021, Hu2022}.
	Studies have also shown sentiment homophily, the tendency for users to interact with others with similar sentiment~\cite{Cinelli2021}.
	These patterns can manifest different forms: pluralism (where individual sentiment is largely independent), consensus (where users converge towards similar sentiment), or polarization (where two or more sentiment clusters emerge)~\cite{Valensise2023}.
	A large scale experiment on Facebook showed that emotional states can spread through this online social network~\cite{Kramer2014}.
	While echo chambers have been documented on other platforms, Reddit has shown less pronounced polarization in contexts such as the 2016 US election and vaccination~\cite{Cinelli2021, DeFrancisciMorales2021}.

	Computational models of social contagion have been applied  to understand the mechanisms behind these emergent patterns of sentiment on social platforms~\cite{Kozitsin2022}.
	These models distinguish between simple contagion, where a single exposure can trigger behavioral change, and complex contagion, requiring multiple exposures~\cite{Guilbeault2018, State2015}.
	Bounded confidence models, which assume that individuals only interact with others holding sufficiently similar opinions, have been used to reproduce observed steady-state patterns of consensus and polarization on social media~\cite{Valensise2023, Baumann2020}.
    In public health, these models have been used to model, for example, the spatial diffusion of vaccine hesitancy~\cite{Haensch2023}, or to couple opinion formation with epidemic spread on contact networks~\cite{Zeng2024}.
    Nevertheless, most studies of online COVID-19 discourse have focused on describing sentiment patterns rather than developing mechanistic models that can explain the underlying dynamics.
	Here we applied a novel dynamic bounded confidence modeling to COVID-19 mitigation discussions on Reddit, improving mechanistic understanding of how sentiment evolves in threaded conversations.

	This study examined discussions on COVID-19 mitigation measures within the Belgian Reddit community
	(r/Belgium) from 1 January 2020 to 30 June 2022, focusing on three key topics: \emph{lockdowns}, \emph{masks}, and \emph{vaccination}.
	Topic modeling and sentiment analysis were used to characterize discussion volume and sentiment dynamics, in pursuit of three goals.
	First, we investigated whether discussion topics exhibited social contagion: a null model for contagion testing revealed no evidence of social contagion in topic initiation.
	Second, we examined whether expressed sentiment displayed homophily: combining interrupted time series analysis with homophily quantification, we found statistically significant sentiment homophily in the reply structure.
	Third, we modeled how this homophily formed under local interactions and long-term trends: we developed a novel stochastic bounded confidence model, the Smooth Latent-Expressed Bounded Confidence (SLEBC) model, that distinguishes between a latent state, and the observable sentiment expressed in their comments.
	The SLEBC model reproduced sentiment distributions and homophily better than two ablations: one that did not distinguish between the two sentiment states, and one that used linear instead of bounded confidence updates.

	\section{Methods}\label{sec:methods}

	\subsection{Data Source and Collection}\label{sec:met-data}
	\begin{figure}
	\centering
	\includegraphics{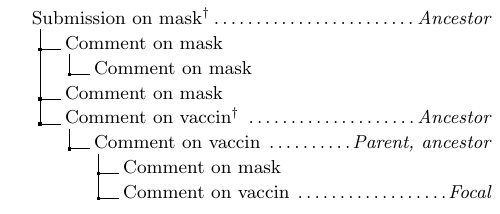}
	\caption{%
		Structure of a Reddit thread.
		The parent and ancestors of the focal comment are shown.
		The initiating posts for the \emph{mask} and \emph{vaccination} discussions are denoted by\ \textsuperscript{\textdagger}\, the others are participating (see Section~\ref{sec:met-activity}).
	}%
	\label{fig:reddit}
	\end{figure}

	Reddit is a social media platform organized into communities called subreddits, denoted by the prefix r/.
	Users operate under fixed pseudonymous usernames and typically share limited personal information.
	Discussions begin when a user posts a submission consisting of a title, optionally accompanied by a hyperlink or text.
	Other users can comment on submissions and reply to specific comments, creating a tree-like conversation structure termed a thread (Figure~\ref{fig:reddit}).
	Throughout this work, the term post refers to both submissions and comments.
	Each comment's parent is the post it replies to, and its ancestors comprise the parent and the ancestors of the parent, always including a single submission at the top level.
	Users can upvote or downvote posts, generating a score reflecting community agreement~\cite{Medvedev2019}.

	Posts from the r/Belgium subreddit created between 1 January 2020 and 30 June 2022 were retrieved from the Pushshift dataset~\cite{Baumgartner2020}, a monthly-updated Reddit archive\footnote{Available at \url{https://academictorrents.com/details/56aa49f9653ba545f48df2e33679f014d2829c10}}.
	This period encompassed the phases before, during, and after the COVID-19 outbreak in Belgium.
	Submissions containing at least one of the keywords listed in Table~\ref{tab:words} in their title, text, or thread were initially selected.
	For each post, the unique identifier, creation timestamp, author, text, score, parent post, and submission title were extracted.
	Markdown tags, URLs, and quotes were removed from all texts.

	\begin{table}[bp]
		\centering
		\begin{tabular}{cccccccc}
			&&&&&&&\\\hline
			corona & virus & covid & mask & masque & lockdown & confin & quarant \\
			curfew & avondklok & couvre-feu & couvre feu & vaccin & vax & jab & booster \\
			prik & piqûre & piqure & pcr & plf & locator form & cst & safe ticket\\\hline
		\end{tabular}
		\caption{Keywords used to select submissions.}%
		\label{tab:words}
	\end{table}

	Topic classification was performed using a multilingual bidirectional encoder representations from transformers (mBERT) model, mbert-corona-tweets-belgium-topics (mbert-ctbt)~\cite{Scott2021}, trained specifically for the Belgian context and capable of processing Dutch, French, German, and English text.
	Although the model identified eight specific measures plus \emph{other measure} and \emph{not applicable} categories, analysis focused on three primary measures: \emph{lockdowns}, \emph{masks}, and \emph{vaccination}.
	These topics were selected for their sustained relevance throughout the pandemic and minimal risk of confusion with non-COVID-19 content.
	Submissions classified as \emph{not applicable} with at least 90\% of their comments similarly classified were excluded.
	Comments labelled \emph{not applicable} or \emph{other} inherited their parent's topic.
	For each topic, the distribution of posts per user was approximated by a power-law distribution \( P(k) \sim k^{-\gamma} \), where \( \gamma \) was determined using maximum likelihood estimation and \( P(k) \) is the probability that a user was author of \( k \) posts~\cite{Clauset2009, Thukral2018}. 

	Language detection was performed using a robustly optimized BERT approach (RoBERTa), xlm-roberta-base-language-detection~\cite{Papariello2024}.
	For English-language posts, sentiment analysis was conducted with twitter-roberta-base-sentiment-latest (roberta-tbsl)~\cite{Loureiro2022}, a classifier used for Reddit content~\cite{Bonifazi2022b, Huang2025}.
	The restriction to English avoided language-specific features in sentiment analysis, while incorporating the bulk of the posts (Section~\ref{sec:met-data}).
	The roberta-tbsl model outputs three scores (0 to 1) for \emph{positive}, \emph{neutral}, and \emph{negative} tonality.
	These were combined into a single continuous sentiment value \( s_k \) per post \( k \) by subtracting the \emph{negative} score from the \emph{positive} score, yielding values from -1 (most negative) to 1 (most positive).
	This continuous representation was employed in sentiment analyses (Sections~\ref{sec:met-sentiment} and~\ref{sec:met-model}).

	\subsection{Temporal Analysis}\label{sec:met-temporal}
	The daily volume of posts related to each topic was calculated and visualized alongside key Belgian policy events, hospitalization counts and administered vaccination doses~\cite{Alleman2021, Sciensano2024}.
	A linear interrupted time series model was fitted to daily post counts, using the event dates in Table~\ref{tab:temporal} as breakpoints~\cite{Bernal2017}.
	The model assumed a linear segment between consecutive breakpoints.
	At each event~\( j \), both the intercept (level change~\( \Delta\beta_{0,j} \)) and the slope (trend change~\( \Delta\beta_{1,j} \)) may change, so yielding a piecewise linear fit with ordinary least squares confidence intervals.
	This allowed us to identify changes in posting volume associated with specific events.
	\ref{app:equations} provides an exact formulation of this model.

	To identify days with unusually negative sentiment, each post's sentiment was weighted by its score, and aggregated per day and per topic.
	A day was marked as a significantly negative day for topic \( x \) if the median weighted sentiment on that day fell below the 0.275 quantile of the overall weighted sentiment distribution for that topic and at least 50 comments were submitted that day.
	These thresholds were chosen to identify one to five significantly negative days per topic.
	Formal definitions of the weighted sentiment aggregates are given in \ref{app:equations}.

	\subsection{Topic Contagion}\label{sec:met-activity}
	To assess whether users were more likely to initiate discussions on a topic after previously participating in one, each post in a thread sharing a common topic was classified as either initiating or participating.
	A post was classified as initiating if none of its ancestors shared the same topic; otherwise, it was participating (Figure~\ref{fig:reddit}).
	The author of an initiating post was designated the initiator, while all others were participants.

	If contagion were present, exposure to a topic through participation increases the probability of later initiating a new discussion on that topic.
	Hence, participations would accumulate at the start of a user's posting sequence.
	Conversely, if topic initiation is independent of prior participation, initiations and participations should be randomly interleaved across the sequence of a user's posts.

	For each user, the number and sequence of initiations and participations per topic were recorded, along with the position in their posting sequence at which the first initiation occurred.
	The observed position was compared against a null model that preserves the number of participations and initiations, but treats each possible ordering as equally likely.
	Under this null model, the probability that the \( i \)'th post is the first initiation has a closed-form combinatorial expression (\ref{app:equations}).
	For each user, the observed position in the sequence where the first initiation occurred was compared to the null model distribution.
	We aggregated this into \( \rho(i) \), the proportion of users with at least one initiation among their first \( i \) discussions.

	\subsection{Sentiment Homophily}\label{sec:met-sentiment}
	The relationship between comment and parent sentiment was examined by comparing each comment's sentiment \( s_k \) to its parent's sentiment \( s_l \).
	All such pairs \( \left( s_k, s_l \right) \) within a topic were collected into a two-dimensional joint histogram with square bins of width \( w_H = 0.05 \), yielding an empirical joint probability distribution \( H \).

	A null model randomly paired comment and parent sentiment by sampling them independently from the observed distributions per topic.
	For comments, this distribution consisted of observed comment sentiment, while the parent distribution additionally included submission sentiment, making the null histogram \( \widetilde{H} \) asymmetric.
	The difference between empirical and null histograms,
	\begin{equation}
		\Delta H =  F_{0.05}\left(H - \widetilde{H}\right), \label{eq:sentdiff}
	\end{equation}
	revealed which sentiment pairs occurred more or less often than expected.
	The function \( F_{0.05} \) sets entries to zero if their \( p \)-value under the null model exceeds 0.05.
	That is, \( \Delta H_{l,m} > 0 \) indicates that the sentiment pair \( (l, m) \) occurred significantly more frequently than expected under independent pairing, and \( \Delta H_{l,m} < 0 \) that it occurred significantly less frequently.
	In the resulting histograms, pluralism appears as uniform patterns, consensus appears as positive mass near a single point on the diagonal, and polarization appears as distinct on-diagonal clusters~\cite{Valensise2023}.

	This two-dimensional histogram was then summarized into a single homophily measure \( h \) by weighting the value of each bin according to its distance from the diagonal \( k = l \),
	\begin{equation}\label{eq:homophily}
		h(\Delta H) = w_H^2\sum_{l,m} \Delta H_{l,m} \left(1 - 2| l - m |\right),
	\end{equation}
	with a summation over all bin midpoints \( l, m \) in \( \Delta H \), and \( w_H^2 = 0.0025 \) is the bin area\footnote{A sensitivity analysis is available in \ref{app:sensitivity}}.
	Positive mass near the diagonal increases \( h \), while off-diagonal mass decreases it.
	Higher \( h \) indicates stronger homophily, while negative values imply heterophily.

	Two contexts served as extensions to comment-parent homophily.
	In each of them, we include more than one (\( n \)) preceding comments.
	First, the ancestral context \( A^n_k \) comprises the \( n \) closest ancestors of comment \( k \).
	Second, the user context \( U^n_k \) includes the parents of comment \( k \) and the \( n-1 \) other comments by the same author on the same topic.
	If users primarily adapted to recently viewed comments, ancestral homophily would dominate.
	If users had inherent sentiment and selectively interacted with similar users, user context homophily would prevail.

	Two-dimensional histograms were created for comment sentiment and mean context sentiment for context sizes one through five.
	Only comments with both ancestral and user contexts of at least size five were analyzed.
	Homophily values per context size were compared to null histograms \( \widetilde{H}^n \) generated by pairing observed sentiment with the mean of \( n \) random observed sentiment.

	\subsection{Sentiment Evolution Model}\label{sec:met-model}
	\begin{figure}[t]
		\centering
		\includegraphics{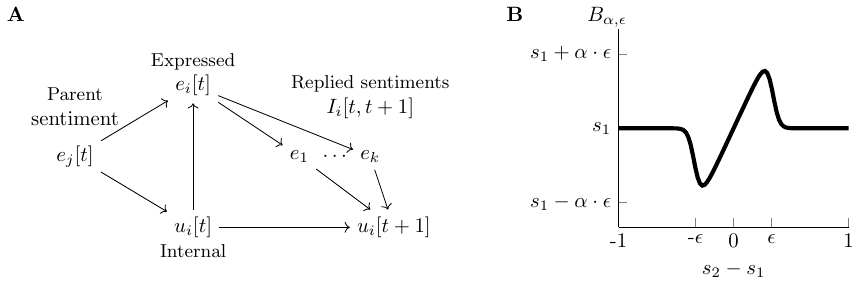}
		\caption{%
			(\textbf{A}) Schematic representation of the Smooth Latent-Expressed Bounded Confidence (SLEBC, Equation~\eqref{eq:model}).
			(\textbf{B}) The change in sentiment after applying \( B_{\alpha, \epsilon} \) (Equation~\eqref{eq:bc}) as a function of the sentiment difference \( s_2 - s_1 \).
		}%
		\label{fig:model}
	\end{figure}

	The Smooth Latent-Expressed Bounded Confidence (SLEBC) model was developed to capture the evolution of a user's sentiment.
	The model includes a latent \( u_i[t] \) and an expressed \( e_i[t] \) sentiment state for each user \( i \).
	The former is an unobserved variable that accumulates sentiment influence across all interactions over time, whereas the latter quantifies the sentiment of a specific comment and is directly observable.
 	This structure is analogous to a state-space model, with \( u_i[t] \) as the latent state and \( e_i[t] \) as a parent-influenced observation.
	The latent state \( u_i[t] \) evolves based on the expressed sentiment of other users, i.e.\ the posts that user \( i \) replies to, and the comments that user \( i \) receives.
	The expressed state \( e_i[t] \) updates according to the user's latent state and the expressed state of the immediate parent comment, allowing a local adaptation.
	Both states evolve stochastically according to a smooth bounded confidence kernel \( B_{\alpha, \epsilon} \), each with its own update strength \( \alpha \), but a shared sentiment distance threshold \( \epsilon \).
	Schematically it is given by Figure~\ref{fig:model}A, and  mathematically by
	\begin{subequations}\label{eq:model}\begin{align}
	 	e_i[t] & \sim \tN{B_{\alpha_{e, i}, \epsilon_i}(u_i[t-1], e_j[t])}{\sigma_{e,i}} \label{eq:model_e}\\
	 	u_i[t] & \sim \tN{\bigotimes_{e_{k}\in I_i[t-1, t]} B_{\alpha_{u,i}, \epsilon_i}(u_i[t-1], e_{k})}{\sigma_{u,i}},\label{eq:model_u}
	\end{align}\end{subequations}
	where \( t-1 \) is the time at which user \( i \) made its previous comment and \( \tN{\mu}{\sigma} \) is a truncated normal distribution with mean \( \mu \) and standard deviation \( \sigma \) and support \( [-1, 1] \).
	In Equation~\eqref{eq:model_e}, \( e_j[t] \) refers to the expressed sentiment of the user \( j \) that user \( i \) is reacting to.
	In Equation~\eqref{eq:model_u}, \(  I_i[t-1, t] \) is the set of sentiment of the parent comment and all replies user \( i \) received between time \( t-1 \) and \( t \), and the operator \( \otimes \) indicates that the bounded confidence update is applied to each such interaction.
	Equation~\eqref{eq:model_e} thus captures how each comment adapts toward the sentiment of its parent, whereas Equation~\eqref{eq:model_u} accumulates the influence of all interactions on the user's latent sentiment trajectory.
	The user-specific parameters \( \alpha_{e,i}, \alpha_{u,i}, \epsilon_i, \sigma_{e,i}, \sigma_{u,i} \) were sampled from a posterior distribution, given the likelihood of the observed expressed sentiment.

	In bounded confidence models, interacting agents align their sentiment (with strength \( \alpha \)), but only if their difference is smaller than a threshold \( \epsilon \), which enables polarized states to emerge~\cite{Hegselmann2002}.
	A smoothing kernel was used to ease calibration,
	\begin{equation}\label{eq:bc}
		B_{\alpha, \epsilon}(s_1, s_2) = s_1 + \frac{\alpha(s_2 - s_1)}{1 + \exp\left(\eta \left({(s_2 - s_1)}^2 - \epsilon^2 \right)\right)},
	\end{equation}
	where  \( s_1 \) is the focal sentiment, \( s_2 \) the sentiment of the other agent, \( \epsilon \) the sentiment difference at which the update is the largest, and \( \alpha \) the strength of the update.
	The shape parameter (fixed at 50) controls how well the smooth kernel approximated the discrete one~\cite{Brooks2024a}.
	Figure~\ref{fig:model}B illustrates the dependence of \( B_{\alpha, \epsilon} \) on the sentiment difference \( s_2 - s_1 \).
	Since \( \eta \) is finite, the maximum sentiment update  remains smaller than \( \alpha \cdot \epsilon \).

	To illustrate the advantage of the bounded confidence kernel \( B_{\alpha, \epsilon} \), it was compared to two alternatives.
	The first uses a linear kernel to update the sentiment in Equation~\eqref{eq:model}~\cite{DeGroot1974}.
	This corresponds to the bounded confidence update \( B_{\alpha, \epsilon} \) in Equation~\eqref{eq:bc} for \( \eta \rightarrow \infty \) and \( \epsilon = 2 \).
	Second, we ablated the latent sentiment state, so that expressed sentiment updated directly against the parent comment and received replies, without a separate latent trajectory (see \ref{app:equations}, Equations~\eqref{eq:linear} and~\eqref{eq:stateless}).

	To fit the SLEBC model and its alternatives, users with at least 40 comments per topic were selected.
	For each user, the sentiment of its comments and the posts it interacted with was extracted.
	The posterior distribution of the parameters was obtained using a Hamiltonian Monte Carlo method, with an exponential prior with mean 0.5 for each \( \sigma \) and each \( \alpha \), and a uniform prior on \( [0, 2] \) for \( \epsilon \)~\cite{Ge2018}.
	The latter prior was chosen such that each meaningful value of \( \epsilon \) was equally likely.
	The prior for the \( \alpha  \)'s was chosen for it to allow high values to capture strong updates, while still limiting the ability to cross the entire sentiment spectrum in a single interaction.
	Results comprised 3000 samples, generated by a no U-turn sampler in six independent chains, each initiated by 250 discarded warm-up samples~\cite{Hoffman2014}.

	To assess the quality of the models, the posterior distribution of predicted sentiment was scored using the Watanabe-Akaike information criterion (WAIC)~\cite{Watanabe2010}.
	A lower value is better, and indicates how well the model can predict data it was not trained on.
	See \ref{app:equations} for details.
	A boxplot synthesizes the homophily values given by \( h \) (Equation~\eqref{eq:homophily}), showing their minimum, maximum, and 0.25, 0.5 and 0.75 quantiles.
	Mann-Whitney \( U \) tests were used to assess the relative strength of latent and expressed update strengths per user, checking whether the probability \( P( \alpha_{e,i} > \alpha_{u,i}) \) is larger than \( P( \alpha_{e,i} < \alpha_{u,i}) \) at the 0.05-level~\cite{Mann1947}.
	The proportion of users for which this is the case is denoted as \( \kappa \).
	Inferred parameters enabled to reconstruct the latent sentiment trajectory of the users over the period of interest.
	Monotonic trends in the population median of inferred latent sentiment were identified by a Hamed-Rao test, a modification of the Mann-Kendall test that accounts for autocorrelation~\cite{Hamed1998}.
	With this test, we checked for each of the subsequent periods delimited by the events in Table~\ref{tab:temporal} if the population median showed such monotonic trend with 0.05 significance.

	\subsection{Ethical Considerations}\label{sec:met-ethics}

	This study was not subject to formal ethics board review, as Belgian law restricts such requirements to medical experiments, animal studies, and dual-use research, consistent with Ghent University's institutional ethics framework~\cite{FOVVVL2004}.
	Data were retrieved from the public subreddit, in line with Reddit's Terms of Service~\cite{RedditSupport2025}.
	Although the raw extract contained identifiers, these were not included in the analytical dataset and were not published; all reported results are aggregated and anonymized.
	This study therefore follows AoIR ethical guidelines for internet research~\cite{franzke2020}.

	\section{Results}\label{sec:results}

	\subsection{Dataset Characteristics}\label{sec:res-dataset}

	\begin{table}[bhp]
    \centering
    \begin{tabular}{rccc}	\hline
			                      & \textbf{Users} & \textbf{Comments} & \textbf{Submissions} \\\hline
			\textbf{Total}        & 28559 & 645280   & 10362 \\\hline
			\textbf{Per language} &       &          & \\
			\emph{EN}            & 24850 & 488468   & 6405 \\
			\emph{NL}            & 9344  & 71940    & 3240 \\
			\emph{FR}            & 696   & 1036     & 145 \\
			\emph{DE}            & 272   & 335      & 4   \\\hline
			\textbf{Per topic}    &       &          & \\
			\emph{Lockdowns}     & 9987  & 94494    & 1009 \\
			\emph{Masks}         & 6824  & 48500    & 437 \\
			\emph{Vaccination}   & 5552  & 41700    & 590 \\\hline
		\end{tabular}
		\caption{Number of users, comments and submissions in the final dataset.}%
    \label{tab:dataset}
	\end{table}

	\begin{figure}[t]
		\centering
		\includegraphics{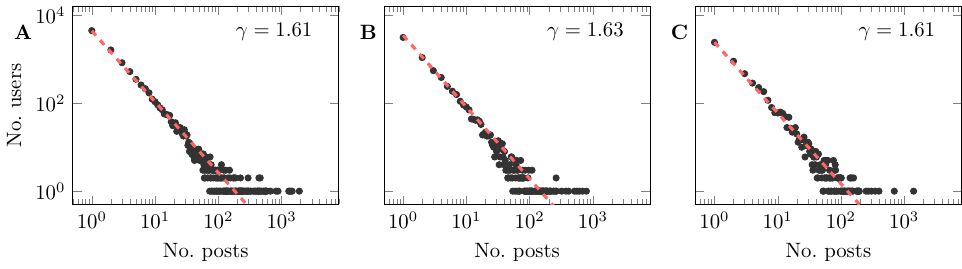}
		\caption{Distribution of posts per user on a double logarithmic scale for (\textbf{A}) \emph{lockdowns}, (\textbf{B}) \emph{masks}, and (\textbf{C}) \emph{vaccination}, and a power-law approximation with exponent \(\gamma\) (dashed, red).}%
		\label{fig:dd}
	\end{figure}

	The final dataset comprised 28,559 unique users who created 645,280 comments and 10,362 submissions (Table~\ref{tab:dataset}).
	English was the predominant language, accounting for approximately 75\% of the posts.
	Dutch posts represented roughly 10\%, while French and German posts were substantially underrepresented with less than 1\% of the posts.
	Topic distribution was relatively balanced, with \emph{lockdowns} being most prevalent (94,494 comments, 1,009 submissions from 9,987 users), followed by \emph{masks} and \emph{vaccination}.

	The distribution of posts per user exhibited a heavy-tailed pattern for all topics (Figure~\ref{fig:dd}), consistent with previous observations on Reddit~\cite{Mancini2022}.
	A small number of users created many posts, while the majority contributed only a single post, limiting the ability to draw conclusions about individual user behavior patterns.
	Power-law approximations yielded similar exponents \( \gamma \) for all topics.

	\subsection{Temporal Analysis}\label{sec:res-temporal}

	\begin{figure}[htp]
		\centering
		\includegraphics{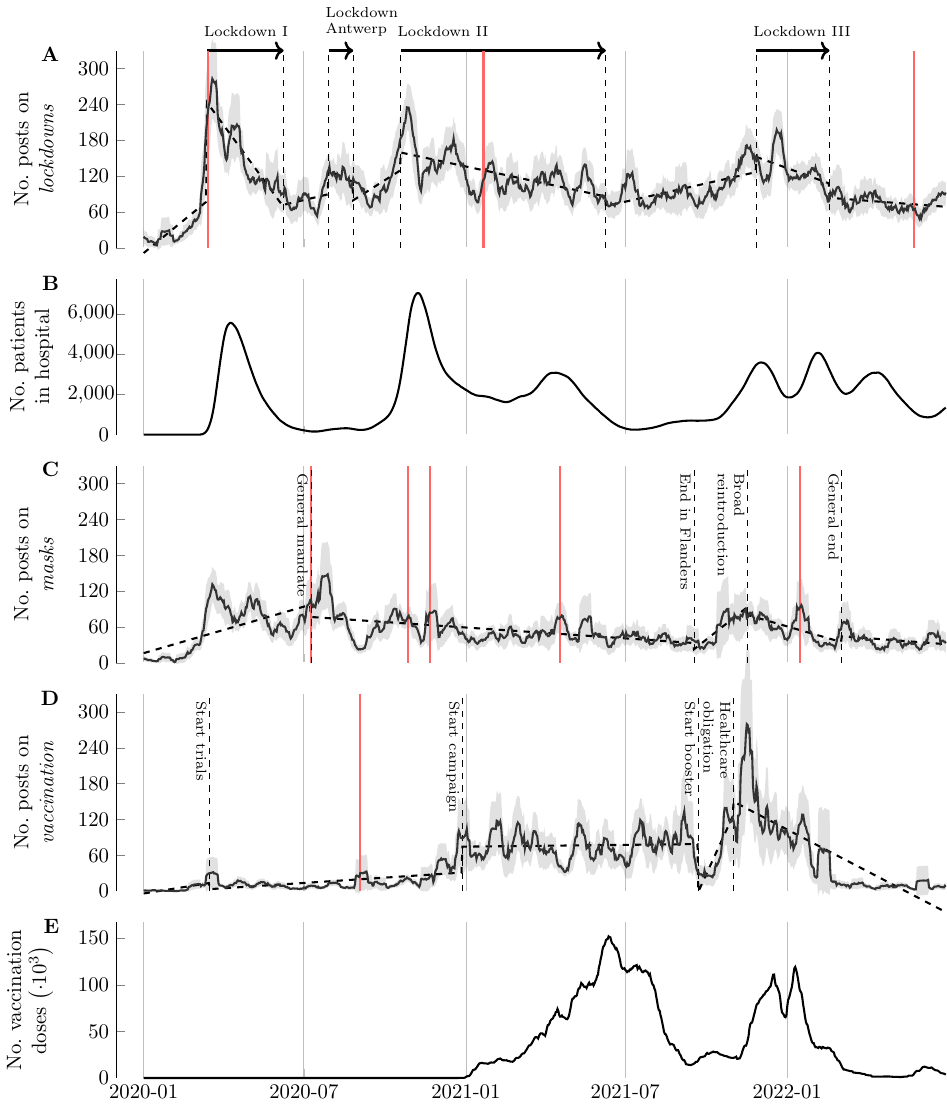}
		\caption{%
			The number of posts per day on r/Belgium (\textbf{A}) \emph{lockdowns},  (\textbf{C}) \emph{masks}, (\textbf{D}) \emph{vaccination}, and (\textbf{B}) the number of  hospitalisations and (\textbf{E}) vaccination doses in Belgium.
			All quantities are shown as their two-week rolling mean.
			Piecewise linear trends are given by dashed lines.
			Red vertical lines mark significantly negative days.
			Important events are denoted by vertical dashed lines or gray regions Table~\ref{tab:temporal}.
		}%
		\label{fig:temporal}
	\end{figure}

	\begin{table}
		\centering
		\begin{tabular}{p{.25\textwidth}lcccc}\hline
			\textbf{Event} & \textbf{Date} & \(\Delta\beta_{0,j}\) & CI 95\% & \(\Delta\beta_{1,j}\) & CI 95\% \\\hline
			\emph{Lockdowns} &&&&&\\
			Lockdown I & 2020-03-13 & \textbf{165.17} & [151.39, 178.96] & \textbf{-3.22} & [-3.53, -2.91] \\
			& 2020-06-08 & 3.82 & [-11.32, 18.96] & \textbf{2.36} & [1.91, 2.81] \\
			Lockdown Antwerp & 2020-07-29 & \textbf{39.85} & [19.77, 59.93] & -1.05 & [-2.14, 0.04] \\
			& 2020-08-26 & \textbf{-30.33} & [-50.72, -9.94] & \textbf{1.64} & [0.56, 2.72] \\
			Lockdown II & 2020-10-19 & \textbf{29.43} & [16.24, 42.62] & \textbf{-1.25} & [-1.63, -0.87] \\
			& 2021-06-09 & \textbf{-15.93} & [-24.62, -7.25] & \textbf{0.65} & [0.57, 0.73] \\
			Lockdown III & 2021-11-27 & \textbf{25.16} & [13.66, 36.67] & \textbf{-0.87} & [-1.08, -0.66] \\
			& 2022-02-18 & \textbf{-23.90} & [-36.02, -11.77] & \textbf{0.42} & [0.20, 0.64] \\
			\hline
			\emph{Masks} &&&&&\\
			General mandate & 2020-07-09 & \textbf{-20.73} & [-27.74, -13.72] & \textbf{-0.53} & [-0.58, -0.47] \\
			End in Flanders & 2021-09-17 & -10.56 & [-21.44, 0.33] & \textbf{1.27} & [0.97, 1.56] \\
			Broad reintroduction & 2021-11-17 & \textbf{-13.69} & [-26.67, -0.72] & \textbf{-1.59} & [-1.91, -1.27] \\
			General end & 2022-03-04 & 10.10 & [-0.63, 20.82] & \textbf{0.31} & [0.15, 0.48] \\
			\hline
			\emph{Vaccination} &&&&&\\
			First trials & 2020-03-16 & -10.36 & [-22.61, 1.90] & -0.14 & [-0.39, 0.11] \\
			Start campaign & 2020-12-28 & \textbf{43.29} & [35.32, 51.26] & \textbf{-0.08} & [-0.13, -0.03] \\
			Start booster & 2021-09-22 & \textbf{-79.62} & [-95.27, -63.96] & \textbf{3.21} & [2.57, 3.86] \\
			Healthcare obligation & 2021-11-19 & \textbf{21.46} & [5.20, 37.73] & \textbf{-4.00} & [-4.64, -3.35] \\
			\hline
		\end{tabular}
		\caption{
			Results of the interrupted time series analysis.
			Change in level \( \Delta\beta_{0,j} \) and in slope \( \Delta\beta_{1,j} \) at the time of each event \( j \) per topic, with their 95\% confidence intervals.
			Significant changes are indicated in boldface.
		}
		\label{tab:its}
	\end{table}

	\begin{table}[htbp]
		\centering
		\begin{tabular}{p{.18\textwidth}lp{.56\textwidth}l}\hline
			\textbf{Event} & \textbf{Date} & \textbf{Description} & \textbf{Ref.} \\\hline
			\emph{Lockdowns}&&&\\
			Lockdown I & 2020-03-13 & Closure of restaurants, pubs, schools, entertainment venues and non-essential stores, part of a European wave of similar national lockdowns. & \cite{BinnenlandseZaken2020},\cite{Hale2021} \\
			& 2020-06-08 & Reopening of culinary, cultural, amusement and religious establishments. & \cite{Wilmes2020} \\
			Lockdown Antwerp & 2020-07-29 & Introduction of a curfew and restriction of four persons per table in bars and restaurants in the Antwerp province. & \cite{Santens2020} \\
			& 2020-08-26 & End of specific restrictions in the province of Antwerp & \cite{Arnoudt2020} \\
			Lockdown II & 2020-10-19 & Closure of restaurants and pubs, reduction of close contacts to one and introduction of a curfew. & \cite{DeCroo2020} \\
			& 2021-06-09 &	Reopening of restaurants, pubs and cultural and entertainment venues. & \cite{Overlegcomite2021} \\
			Lockdown III & 2021-11-27 & Restriction of the number of persons per table and opening hours in restaurant and bars. Prohibition of private and events without seating. & \cite{BinnenlandseZaken2021} \\
			& 2022-02-18 & Restart of public events and unrestricted opening hours for bars and restaurants. & \cite{Overlegcomite2022} \\\hline
			\emph{Masks}&&&\\
			General mandate & 2020-07-09 & Announcement of a mask mandate for commercial, cultural and religious venues, following Spain and France. & \cite{TT2020},\cite{Hale2021} \\
			End in Flanders & 2021-09-17 & Announcement of reduction of federal mandate to public transport, big events, close contact professions and healthcare. In practice only happened in the region of Flanders. & \cite{Overlegcomite2021a} \\
			Broad reintroduction & 2021-11-17 & Announcement of extension of the federal mask mandate to public and commercial venues.  & \cite{Overlegcomite2021b} \\
			General end & 2022-03-04 & Announcement of mask obligation limited to public transport and healthcare from 7th of March. & \cite{BinnenlandseZaken2022} \\\hline
			\emph{Vaccination}&&&\\
			First trials & 2020-03-16 & Clinical testing on human starts. & \cite{ThanhLe2020} \\
			Start campaign & 2020-12-28 & First vaccination in Belgium. & \cite{Maerevoet2020} \\
			Start booster & 2021-09-22 & Decision to allow third vaccinations for retirement home inhabitants. & \cite{Arnoudt2021} \\
			Healthcare obligation & 2021-11-19 & Federal government reaches agreement about obligation of vaccination in healthcare. & \cite{Paelinck2021} \\\hline
		\end{tabular}
		\caption{Description of the events shown in Figure~\ref{fig:temporal}~\cite{Alleman2021}. Except where explicitly stated, all measures were imposed by the federal government and hence applicable to the whole country.}%
		\label{tab:temporal}
	\end{table}

	Figure~\ref{fig:temporal} shows posting volume over time for the three topics, alongside key events and epidemiological indicators.
	It gives an overview of posting behavior over the period of the pandemic, and associated events.
	The largest peaks in number of posts coincided with the external events listed in Table~\ref{tab:temporal}.
	For example, the start of the first lockdown coincided with the single most active day across all topics, with a rolling mean of 282.79 posts on 19 March 2020.
	Smaller local maxima appeared during the Antwerp lockdown and at the start of the second national one.
	All events coincided with significant changes in post volume.
	The piecewise linear trends overlaid on Figure~\ref{fig:temporal} (dashed lines) make these structural changes visible, with the slope and level shifts quantified in Table~\ref{tab:its}.
	After these peaks, the volume of posts decreased and stabilized.
	Three days showed a significantly negative sentiment: at the start of the first lockdown,  in the middle of the second national one, and in February 2022, after the end of most mitigation measures.
	During the first one, negative comments were aimed at ``lockdown parties'', social gatherings just before the restrictions came into effect.

	Discussions on \emph{masks} started at the start of the COVID-19 outbreak in Belgium, four months before the announcement of a general mandate (9 July 2020), which itself did coincide with a small increase in posts, and a significantly negative day.
	During the latter, negative comments criticized the late decision of the government to introduce the mandate, and people not following it.
	Activity on \emph{mask}-related discussions fell after the first lockdown but rose again during the second mandate period (November 2021 -- March 2022), which included another significantly negative day.
	Another three significantly negative days occurred during the first mandate.
	All events were associated with a significant change in trend, and the two moments the mandate was expanded, did as well with a significant change in level (Table~\ref{tab:its}).
	Post volume generally increased before mask mandates were tightened, and started decreasing afterwards.

	Similar to the lockdown-related events, major vaccine-related events or news corresponded well with the volume of \emph{vaccination} posts on Reddit.
	Broad discussions did not start until the run-up to the vaccination campaign, with the only non-significant change in level or volume occurring at the start of the trials (Table~\ref{tab:its}).
	The announcement that vaccination would be mandatory for healthcare personnel preceded the month with the highest posting activity, culminating in a rolling mean peak of 279.64 posts on 16 November 2021.
	The single significantly negative day, 3 September 2020, fell during the trials.
	Comments reacted to an article stating that 30\% of the Belgian population would refuse a COVID-19 vaccine~\cite{Galindo2020}.
	The negative reactions targeted both refusers and the ``rushed'' development of the new mRNA technology.

	\subsection{Topic Contagion}\label{sec:res-activity}

	\begin{figure}[t]
		\centering
		\includegraphics{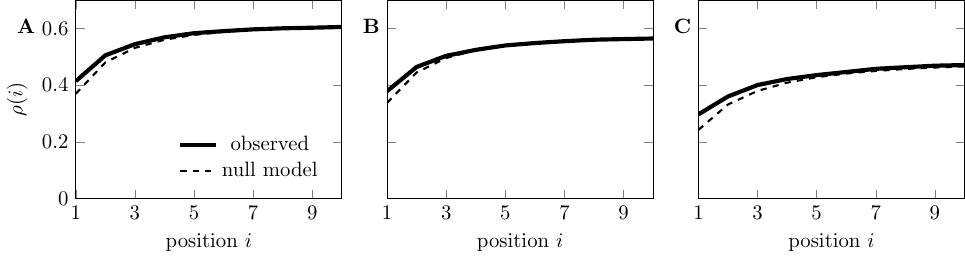}
		\caption{%
			Proportion of users \( \rho(i) \) that has at least one initiated post in the first \( i \) discussions for the three topics, (\textbf{A}) \emph{lockdowns}, (\textbf{B}) \emph{masks}, and (\textbf{C}) \emph{vaccination}.
			The full line represents the observed proportion, the dashed line the expected proportion under the null model.
		}%
		\label{fig:ecdfs}
	\end{figure}

	We test whether prior participation in a discussion on a topic increases the probability of subsequently initiating a new one, which serves as an indication of social contagion.
	Under the null hypothesis (no contagion), initiations and participations are ordered uniformly at random.
	Under contagion, on the other hand, participation would increase the probability of later initiating a new discussion, causing participations to accumulate earlier in users' sequences.
	Figure~\ref{fig:ecdfs} shows the proportion of users with at least one initiation among their first \( i \) discussions, \( \rho(i) \).
	The solid line represents the observed cumulative proportion, and the dashed line is the expected proportion under the null model.
	If contagion were present, \( \rho(i) \) would lie \emph{below} the null for small \( i \), reflecting an accumulation of participations before the first initiation.
	For all topics, the observed curve lies above the null at low values of \( i \), indicating initiations occurred earlier in posting sequences than expected.
	This contradicts the contagion hypothesis: initiations were more likely near the start of sequences.
	Consequently, users who participated in a discussion were not more likely to subsequently initiate one on the same topic.
	No evidence of social contagion was found for discussion initiation on the three mitigation topics within r/Belgium, neither simple nor complex.
	As \( \rho(i) \) converges, it approaches the proportion of users with at least one initiation: 0.612 for \emph{lockdowns}, 0.572 for \emph{masks}, and 0.481 for \emph{vaccination}.
	Hence, not only did discussions on vaccination appear later than those on the other topics, they were initiated by a relatively smaller fraction of the user base

	\subsection{Sentiment Homophily}\label{sec:res-sentiment}

	\begin{figure}[htp]
		\centering
		\includegraphics{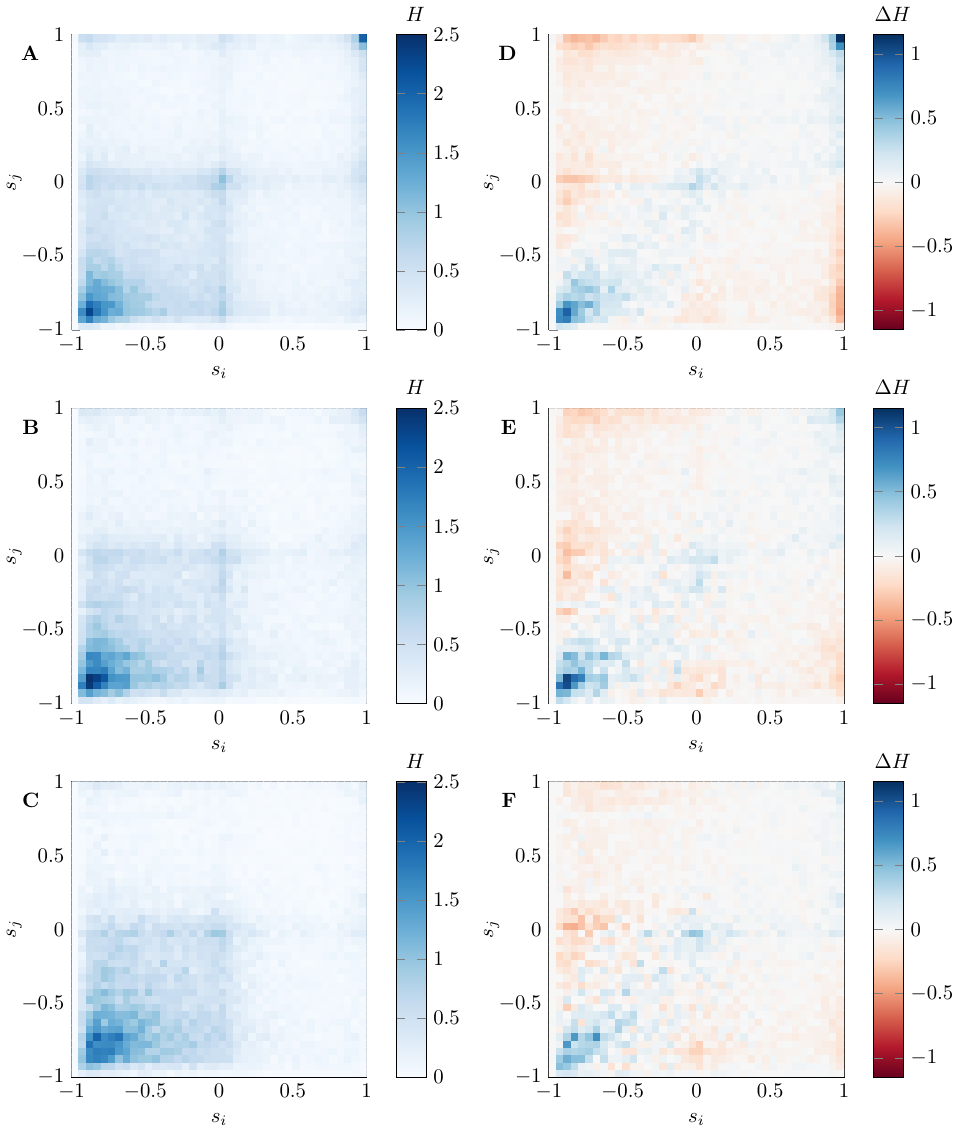}
		\caption{%
			Normalized histograms of (\textbf{A},\textbf{B},\textbf{C}) observed sentiment interactions, and (\textbf{D},\textbf{E},\textbf{F}) the difference with the null model for the topics (\textbf{A},\textbf{D}) \emph{lockdowns}, (\textbf{B},\textbf{E}) \emph{masks}, and (\textbf{C},\textbf{F}) \emph{vaccination}.
		}%
		\label{fig:sent_hm}
	\end{figure}

	We examined the relationship between the sentiment of a comment and that of its parent, looking for correlations between these, namely homophily.
	Figure~\ref{fig:sent_hm} presents heatmaps of comment-parent sentiment pairs \( H \) and \( \Delta H \) for each topic.
	Across all topics, the center of mass of observed interactions \( H \) lay in the fourth quadrant, with a global maximum near (-1,-1), indicating most posts expressed negative sentiment.
	For \emph{lockdowns}, another local maximum appears in the first quadrant corner.
	Discussions around \emph{masks} and \emph{vaccination} exhibited consensus (predominantly negative), while those on \emph{lockdowns} showed polarization (negative replies to negative, positive to positive)~\cite{Valensise2023}.

	Both consensus and polarization fall under the broader term homophily, the tendency to interact with others expressing similar sentiment.
	Homophily is expected when the underlying distribution is narrow, offering few interaction alternatives.
	However, the heatmaps of \( \Delta H \) in Figure~\ref{fig:sent_hm} demonstrate homophily for all topics, irrespective of sentiment distribution shape.
	Positive values center around the diagonal, indicating more interactions between similar sentiment than expected.
	Negative values appear in the second and third quadrants, and in regions representing interactions between negative and neutral sentiment.
	Discussions on \emph{lockdowns} were most polarized, with clear peaks in the equally-signed corners.
	The discrepancy between observed and expected values reveals greater polarization than \( H \) alone suggested for \emph{masks}, with positive values of \( \widetilde{H} \) near (1,1).
	The homophily measure \( h \) (Equation~\eqref{eq:homophily}) is 0.246, 0.212, and 0.144 for \emph{lockdowns}, \emph{masks}, and \emph{vaccination}, respectively. 
	As Figure~\ref{fig:sent_hm} suggests, homophily is most pronounced for lockdowns, and least for vaccination.
	This mirrors the pattern of Sections~\ref{sec:res-temporal} and~\ref{sec:res-activity} where lockdowns and vaccination produce the most differing results, with masks showing more similarities to lockdowns than to vaccination.

	\begin{figure}[t]
		\centering
		\includegraphics{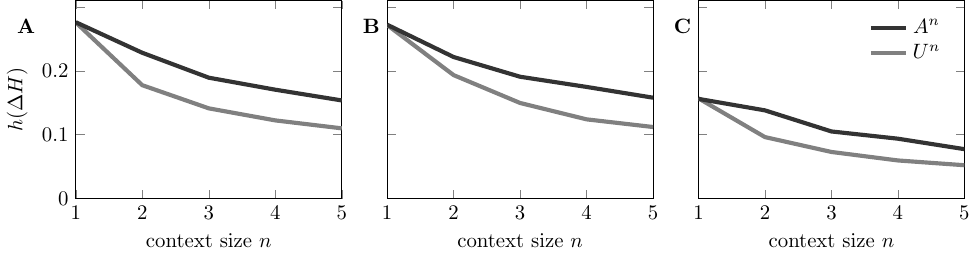}
		\caption{%
			Measure of homophily \( h \) between the sentiment of a comment and its context size \( n \) ancestral (black) or user (gray) context for the topics (\textbf{A}) \emph{lockdowns}, (\textbf{B}) \emph{masks}, and (\textbf{C}) \emph{vaccination}.
		}%
		\label{fig:homo}
	\end{figure}

	We examined which context type exhibited stronger homophily: ancestral \( A^n_k \) or user \( U^n_k \).
	The former comprises posts the focal user most recently read, while the latter comprises posts with which the user most recently interacted.
	For this, a subsample of comments allowing a sufficient size for both contexts was considered (9067, 5372, and 4893 comments for \emph{lockdowns}, \emph{masks}, and \emph{vaccination}, respectively).
	Figure~\ref{fig:homo} presents the homophily measure \( h \) (Equation~\eqref{eq:homophily}) for all three topics and both context types as a function of context size \( n \), the number of preceding posts that are taken into account.
	With \( h \) being 0.276 (\emph{lockdowns}), 0.272 (\emph{masks}), and 0.156 (\emph{vaccination}) for \( n=1 \), homophily in this subset is more pronounced, but the relative ordering of topics persists.
	The homophilic effect decreased with increasing context size \( n \), regardless of type.
	Homophily within the ancestral context consistently exceeded that in the user context for all \( n \)
	Comment sentiment was thus more strongly related to the immediate posting environment than to sentiment that previously prompted replies from the focal user.
	Comments on \emph{vaccination} exhibited less homophily than the other topics.

	\subsection{Sentiment Evolution Model}\label{sec:res-model}

	\begin{table}[bhp]
    	\centering
		\begin{tabular}{rcc}\hline
			\textbf{Topic}     & \textbf{Users} & \textbf{Comments} \\\hline
			\emph{Lockdowns}   & 209   & 16762 \\
			\emph{Masks}       & 101   & 4277 \\
			\emph{Vaccination} & 98    & 3169 \\\hline
		\end{tabular}

		\caption{Number of users and comments used to fit the SLEBC model}%
    \label{tab:modeldata}
	\end{table}

	\begin{table}[bhp]
		\centering
		\begin{tabular}{lccc}
			\hline
			\textbf{Model} & \( WAIC \) & \( \Delta h \) & [95\% CrI] \\\hline
			\emph{Lockdowns} & & \\
			\( L_\alpha \) (Equation~\eqref{eq:linear}) & -21.11 & -0.0973 & [-0.1084, -0.0720] \\
			\( \bar{e} \) (Equation~\eqref{eq:stateless}) & 692 & -0.0702 & [-0.1078, -0.0441] \\
			SLEBC (Equation~\eqref{eq:model}) & -28.5 & -0.0273 & [-0.0511, 0.0015] \\\hline
			\emph{Masks} & & \\
			\( L_\alpha \) (Equation~\eqref{eq:linear}) & -17.4 & -0.1317 & [-0.1651, -0.0972] \\
		    \( \bar{e} \) (Equation~\eqref{eq:stateless}) & 169 & -0.0759 & [-0.1534, -0.0276] \\
			SLEBC (Equation~\eqref{eq:model}) & -18.4 & -0.0671 & [-0.0816, 0.0103] \\\hline
			\emph{Vaccination} & & \\
			\( L_\alpha \) (Equation~\eqref{eq:linear}) & -20.6 & -0.1032 & [-0.1393, -0.0732] \\
		    \( \bar{e} \) (Equation~\eqref{eq:stateless}) & 6.81 & -0.0672 & [-0.1395, -0.0410] \\
			SLEBC (Equation~\eqref{eq:model}) & -21.2 & -0.0337 & [-0.0829, 0.0046] \\\hline
		\end{tabular}
		\caption{%
			Model comparison showing Watanabe-Akaike information criterion (\(WAIC\)) and homophily difference \( \Delta h \) (with 95\% credible interval) for the SLEBC model and its two alternatives, the linear model \( L_\alpha \) and the stateless model \( \bar{e} \).
		}%
		\label{tab:modelcomp}
	\end{table}

	The smooth latent-expressed bounded confidence (SLEBC) model (Equation~\eqref{eq:model}) was set up to give mechanistic insights into the results of Section~\ref{sec:res-sentiment}.
	To fit the parameters of this model and its alternatives, we considered 209, 101, and 98 users for \emph{lockdowns}, \emph{masks}, and \emph{vaccination}, respectively, each with at least 40 comments (Table~\ref{tab:modeldata}).
	These users were the most active and thus yielded sufficient data, though the homophily within this population differed from the complete set (relative difference of -10.5\% for \emph{lockdowns}, +12.5\% for \emph{masks} and +13.7\% for vaccination).
	Consequently, the results might not apply to the more casual users.

	Table~\ref{tab:modelcomp} compares the performance of the three models, with lower WAIC and \( |\Delta h| \) indicating better model fit.
	All models systematically underestimated homophily, as indicated by negative \( \Delta h \), with only SLEBC having 95\% credible intervals that encompass zero.
	The SLEBC model (Equation~\eqref{eq:model}) outperformed both alternatives across all topics, achieving the lowest WAIC values and homophily deficits \( \Delta h \) closest to zero.
	The two alternatives each have their weaknesses.
	The linear model (Equation~\eqref{eq:linear}) achieved comparable predictive power to SLEBC, but substantially underestimated homophily.
	The stateless model (Equation~\eqref{eq:stateless}), which ablated the latent state, achieved a worse fit, but more accurately predicted homophily.
	These results suggest the bounded confidence kernel \( B_{\alpha, \epsilon} \) is better suited to model the sentiment homophily, with the model lacking it underestimating \( h \).
	Moreover, the latent state improves capturing the sentiment distribution over time, by accounting for fluctuating expressed sentiment.

	\begin{figure}[t]
		\centering
		\includegraphics{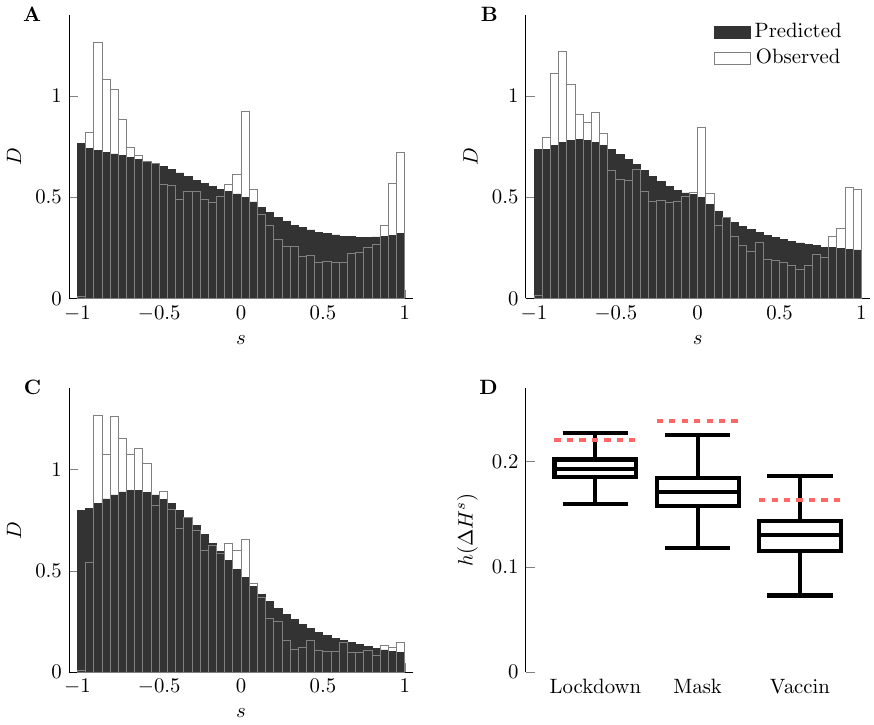}
		\caption{%
			Outcomes of the fitted posterior of the SLEBC model.
			Normalized distributions \( D \) of the predicted sentiment for the topics (\textbf{A}) \emph{lockdowns}, (\textbf{B}) \emph{masks}, and (\textbf{C}) \emph{vaccination} in the SLEBC model, compared to the observed sentiment.
			(\textbf{D}) The predicted values of \( h \) (Equation~\eqref{eq:homophily}) by the SLEBC model, with minimum, maximum and 0.25, 0.5 and 0.75 quantiles.
			The dashed red line represents the observed value.
		}%
		\label{fig:model_hists}
	\end{figure}

	Figure~\ref{fig:model_hists} compares per topic the posterior sentiment distributions and homophily values obtained from the SLEBC model to those observed in the data.
	The SLEBC model predicted sentiment distributions that placed most mass near the negative extreme, while spreading more evenly over their support than observed distributions and lacking their tri-modal structure (Figure~\ref{fig:model_hists}A-C).
	In particular, the local maximum around zero (neutral sentiment) present in the data was not captured.
	On the positive extreme, both observed and predicted distributions exhibited a local maximum for \emph{lockdowns}, though this peak was substantially more pronounced in the data.
	Applying the homophily analysis from Section~\ref{sec:met-sentiment} to the predictions showed that the SLEBC model generated homophily (Figure~\ref{fig:model_hists}D), but underestimated it for all topics, with median values below the observed ones.

	\begin{table}[bhp]
		\centering
		\begin{tabular}{r|ccccccc}
			\textbf{Topic} & \( \epsilon \) & 95\% QI & \( \alpha_u \) & 95\% QI   & \( \alpha_e \) & 95\% QI & \( \kappa \)  \\\hline
			\emph{Lockdowns} & \( 0.973 \) & \( [0.367, 1.566] \) & \( 0.414 \) & \( [0.126, 0.706] \) & \( 0.871 \) & \( [0.269, 3.694] \) & 0.742 \\
			\emph{Masks} & \( 0.874 \) & \( [0.372, 1.510] \) & \( 0.417 \) & \( [0.238, 0.609] \) & \( 0.656 \) & \( [0.265, 1.933] \) & 0.703 \\
			\emph{Vaccination}  & \( 0.971 \) & \( [0.445, 1.484] \) & \( 0.425 \) & \( [0.139, 0.788] \) & \( 0.508 \) & \( [0.234, 1.389] \) & 0.510 \\
		\end{tabular}

		\caption{%
			The population mean values of \( \alpha_u \) and \( \alpha_e \) with their central 95\% quantile intervals (95 \% QI) in the SLEBC model (Equation~\eqref{eq:model}) sampled from the posterior distribution,
			together with \( \kappa \), the proportion of users for which \( P(\alpha_{u, i} < \alpha_{e, i}) > P(\alpha_{u, i} > \alpha_{e, i}) \) with \( p \)-value 0.05.
		}%
    	\label{tab:alphas}
	\end{table}

	Table~\ref{tab:alphas} presents a quantitative summary of update strength parameters \( \alpha_u \) and \( \alpha_e \) in the SLEBC model (Equations~\eqref{eq:bc} and~\eqref{eq:model}).
	For all topics and most users, expressed sentiment was more strongly influenced by parent comment sentiment than latent sentiment was by other interactions, as \( \kappa \) exceeded one half, although only marginally for \emph{vaccination}.
	Adaptation to parent comment sentiment therefore constituted an important mechanism behind the observed homophily, consistent with findings in Section~\ref{sec:res-sentiment}.
	The latent update strength \( \alpha_{u} \) varied less across topics and users within a topic, as shown by its lower standard deviation, indicating that the degree of expressed sentiment adaptation to the parent depended more on user-specific behavior.

	The mean sentiment threshold \( \epsilon \) is lower for masks than the other two topics, while the quantile interval for vaccination is most narrow.
	For all topics, the mean value is just under half of the maximum meaningful value, two.
	The quantile intervals indicate that only a small number of users have very low or very high ones.
	The SLEBC model thus infers that users were moderately open to influence: they adapted their sentiment toward that of others unless opinions diverged by more than roughly half the scale, for example between negative and neutral.
	In contrast, only a limited number of users displayed no sentiment adaptation (\( \epsilon \approx 0 \)), or indiscriminate adaptation (\( \epsilon \approx 2 \)).

	\begin{figure}[tbp]
		\centering
		\includegraphics{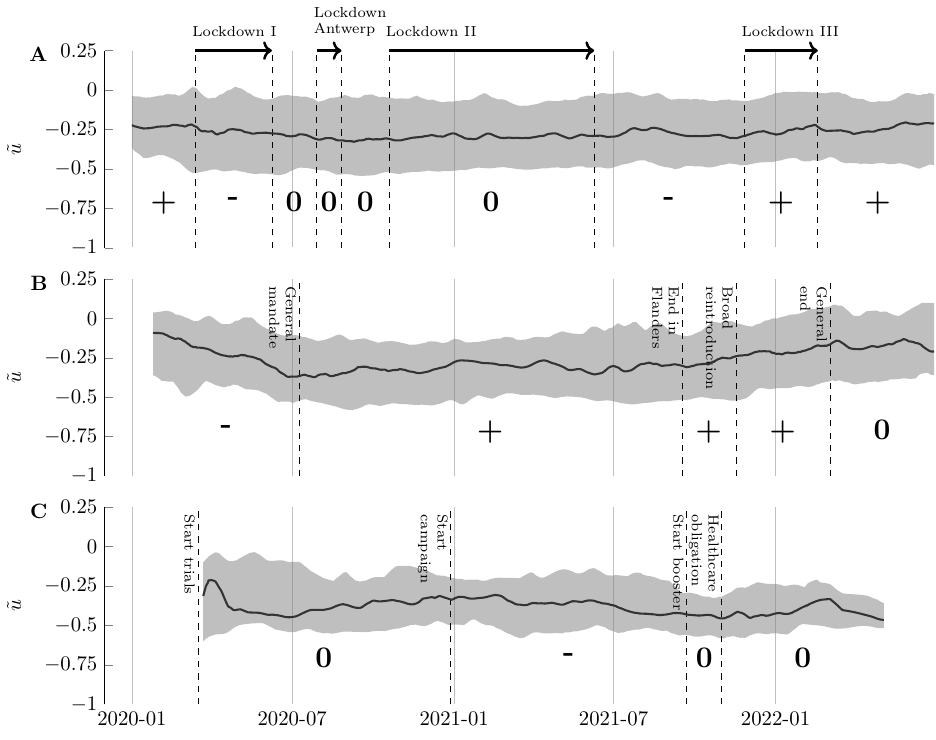}
		\caption{%
			Median latent sentiment \( \tilde{u} \) of users over time (full black curve) for the topics (\textbf{A}) \emph{lockdowns}, (\textbf{B}) \emph{masks}, and (\textbf{C}) \emph{vaccination}.
			The area between the first and third quartile is shaded.
			Hamed-Rao monotonic test result are given per period delimited by the events in Table~\ref{tab:temporal} (+: increasing; -: decreasing; 0: no significant trend).
		}%
		\label{fig:latent}
	\end{figure}

	A key feature of the SLEBC model is its ability to separate the influence of a user's latent sentiment from that of the parent comment on the expressed sentiment.
	For assessing sentiment toward a mitigation measure, the former is particularly informative, as it captures the less noisy, longterm trends.
	Figure~\ref{fig:latent} presents the evolution of the inferred latent state of the SLEBC model, \( u \) for the studied population.
	For \emph{masks}, median latent sentiment declined during the first half of 2020 before gradually increasing.
	For \emph{lockdowns}, latent sentiment remained stably negative throughout the study period.
	The Hamed-Rao test identified both increasing and decreasing trends at the start and end of the considered period, however the absolute change was small.
	This is the case for \emph{vaccination} as well, although sentiment initially showed a short positive tendency around the start of the trials.

	\section{Discussion}\label{sec:discussion}

	\subsection{Principal Findings}\label{sec:disc-principal}

	This study analyzed discussions on COVID-19 mitigation measures on the Belgian Reddit community (r/Belgium) from January 2020 to June 2022, examining how discussion volume and sentiment evolved around \emph{lockdowns}, \emph{masks}, and \emph{vaccination}.
	We found that discussion volume was primarily associated with external events such as policy announcements and media coverage rather than by social contagion within the platform.
	Peaks in posting activity aligned with major policy changes, including the first lockdown and the announcement of mandatory vaccination for healthcare personnel, suggesting that traditional media and official communications may play an important role in shaping health discourse in r/Belgium, as has been confirmed on Twitter~\cite{Kurten2021,Kobayashi2022a} and other Reddit communities~\cite{Gozzi2020}.
	This contrasts with observations for other online behaviors where complex social contagion mechanisms have been documented~\cite{State2015}.

	Although topic-level contagion was not significant, sentiment exhibited homophily, with users adapting their sentiment to match that of the parent comment.
	The observed homophily measure \( h \) ranged from 0.144 to 0.246 across the three topics, confirming patterns previously documented on Reddit~\cite{Cinelli2021, Bonifazi2022b}. 
	Our analysis suggested that this homophily operated primarily through immediate conversational adaptation rather than through selective interaction with like-minded users.
	To capture this dynamic, we developed the SLEBC model, which distinguishes between a user's unobserved latent sentiment and the observed sentiment expressed in comments.
	The model successfully reproduced observed patterns and revealed that most users (51-74\% depending on topic) adjusted their expressed sentiment more strongly to match that of the parent than they updated their latent sentiment states.

	The SLEBC model, building on bounded confidence modeling~\cite{Hegselmann2002}, outperformed linear models, indicating that sentiment alignment occurred only when users encountered other users with sufficiently similar sentiment.
	Our results expand on the existing literature on using the bounded confidence model in steady-state opinion distributions on social media~\cite{Valensise2023}, and spatial patterns of vaccine hesitancy~\cite{Haensch2023}.
	Nevertheless, a linear model explicitly using social network structure has been successfully calibrated on Italian vaccine hesitancy survey data, with the authors naming the bounded confidence model as a possible extension~\cite{Ancona2022}.
	By also incorporating a latent sentiment state, the SLEBC model corrected for the fluctuating nature of the expressed sentiment, and better inferred the observed distribution of sentiment over time.
	This latent state may better represent a user's underlying sentiment trajectory than their fluctuating expressions alone, making it a stronger signal for relating online discourse to behavioral indicators.

	Discussions on different mitigation measures showed distinct patterns on r/Belgium, similarly to Twitter~\cite{Scott2021}.
	Discussions on \emph{lockdowns} dominated in volume and emerged when restrictions were first imposed.
	This aligns with the increase in the volume of COVID-19 related posts observed in other country-related subreddits during a lockdown period~\cite{Hu2022}.
	\emph{Masks} were discussed from the pandemic's onset, before any official mandate.
	\emph{Vaccination}, by contrast, did not attract broad discussion until the run-up to the campaign.
	A possible explanation for the temporal differences is that mask wearing was a personal choice before the official mandate, whereas vaccination was impossible before the government launched the campaign.

	The three topics also differed on sentiment.
	Discussions on \emph{lockdowns} showed stronger polarization, with distinct positive and negative sentiment clusters.
	In contrast, \emph{masks} and \emph{vaccination} both exhibited  predominantly negative sentiment, though \emph{vaccination} showed weaker homophily (\( h = 0.144 \)) than \emph{masks} (\( h = 0.212 \)).
	This is mirrored in the inferred parameters of the SLEBC model.
	For \emph{lockdowns} the mean external update strength is strongest, followed by \emph{masks} and \emph{vaccination}.
	In contrast, population distributions of sentiment thresholds and latent update strengths showed more similarities over the topics.
	Aforementioned differences underscore the importance of analyzing mitigation topics separately rather than aggregating them into a single ``COVID-19 sentiment'' variable.

    Existing coupled epidemiological-social models may implicitly treat social media activity as a direct proxy for the infection dynamics, whereas our results indicate that a communication layer should be considered as well~\cite{Bedson2021}.
    In particular, modeling of discussion volume may benefit from an intermediate layer representing official communications and traditional media such as the OxCGRT stringency index~\cite{Hale2021} or GDELT dataset~\cite{Alipour2024}.
	Users can then react to activity in this layer, allowing for an increase in discussion volume without within platform contagion.
	Sentiment dynamics, on the other hand, are better represented by complex contagion models such as bounded confidence kernels rather than by simple ones.
	The methodological framework of combining topic modeling, sentiment analysis, null model comparisons, and mechanistic modeling, provides a template for digital health surveillance deployment during health emergencies, with established platforms like Reddit offering advantages as data infrastructure and analysis pipelines can be prepared in advance.

	\subsection{Limitations and Future Work}\label{sec:disc-limitations}
	This study has several limitations.
	The r/Belgium community represents a self-selected, non-representative sample of the Belgian population.
	Reddit users tend to belong to a specific demographic, often young and male~\cite{DeFrancisciMorales2021}.
	This effect intersects with our focus on English-language posts (approximately 75\% of the dataset).
	However, because English acts as a lingua franca on r/Belgium (and the majority of other Reddit communities) that bridges the country's distinct linguistic communities, restricting the analysis to English avoids introducing confounding variables from isolated subgroups that communicate exclusively in Dutch, French, or German.
	Moreover, SLEBC parameters were inferred from highly active users, whose homophily values differed from the full sample.
	Inferred parameters and latent trajectories may therefore not represent the sentiment states of casual users.
	Nevertheless, qualitative findings such as the adaptation of expressed sentiment to the parent post are furthermore supported by the full dataset analysis in Section~\ref{sec:res-sentiment}.
	Still, our findings represent platform- and user-specific dynamics that may not generalize, as highly active, English-proficient Reddit users may differ systematically in demographics, political orientation, or health attitudes from other communities.
	Applying our approach across different communities and platforms would help clarify how specific platform architectures influence sentiment homophily~\cite{Pierri2023, Yurtsever2023, Kobayashi2022a}.

	Sentiment analysis captures linguistic tone rather than explicit policy support, introducing ambiguity where negative sentiment could reflect criticism of either insufficient or excessive measures.
	The rapid increase in ability and availability of large language models can enhance our methodology by extracting more nuanced signals such as support for public health measures and willingness to adhere.
	The inferred latent states are unobservable mathematical constructs.
	While our model successfully reconstructs latent sentiment, it remains to be established whether these trajectories serve as reliable proxies for true offline attitudes.
	Future work could address this by relating the inferred latent states with behavioral surveys or adherence metrics.

	\subsection{Conclusions}\label{sec:disc-conclusion}

	Our study showed that discussions on COVID-19 mitigation measures on Reddit's r/Belgium were not driven by social contagion within the platform, but aligned with external events and traditional media.
	In contrast, sentiment dynamics were shaped by within-thread interactions.
	Users exhibited sentiment homophily by adapting their expressed sentiment to match that of preceding comments in the thread.
	The SLEBC model captures this by distinguishing between an unobserved latent sentiment state and sentiment expressed in comments, suggesting that expressed sentiment adapts strongly to align with that of the parent comment, and therefore may poorly reflect users' underlying sentiment.

	These findings have implications for Reddit-based digital health surveillance and epidemic modeling that uses social media signals.
	First, rather than assuming purely endogenous social contagion, models benefit from an intermediate communication layer of external official communications and traditional media as the driver of topic emergence.
	Second, when integrating social sentiment as a proxy for public adherence, models should account for interaction structure.
	In COVID-19 mitigation discussions on r/Belgium, expressed sentiment adapted primarily to the immediate parent comment rather than to a user's broader interaction history, and this adaptation was better captured by a bounded confidence kernel than by a linear update, with the latter substantially underestimating observed homophily.
	Raw expressed sentiment is therefore a noisy, perturbed signal.
	The inferred latent trajectory is more stable and, if validated against behavioral outcomes, a stronger candidate for model input.
	Finally, mitigation measures exhibited distinct patterns in our research and should not be collapsed into a single ``COVID-19 sentiment'' variable, which would obscure relevant differences.

	\section*{Funding}
	This work was supported by the Research Foundation --- Flanders (FWO) (grant numbers G0G0122N, W001625N, L.E.C.R.\ and J.M.B.), Special Research Fund (grant number 2024/01/709, L.E.C.R.), Atlantic Coast Center for Infectious Disease Dynamics and Analytics - ACCIDDA (CDC CFA grant NU38FT000012, T.W.A.) and National Institutes of Health grant number 5R24GM153920‑02 (T.W.A).
	Computational infrastructure was provided by the Flemish Supercomputer Center (VSC).
	Published with the support of the University Foundation of Belgium.
	The organisations did not in any way influence the design, execution or results of this work.

	\section*{Conflicts of Interest}
	None declared.

	\section*{Data Availability}
	Instructions on how to obtain the used dataset (\url{https://academictorrents.com/details/56aa49f9653ba545f48df2e33679f014d2829c10}~\cite{Baumgartner2020}) and reproduce the findings are available in the GitHub repository at \url{https://github.com/TimVWese/covid-reddit-belgium}.
	According to the Reddit Terms of Service, their data is freely usable for non‑profit purposes~\cite{RedditSupport2025}.

	\section*{Authors' Contributions}
	Tim Van Wesemael:\@ Conceptualization, Methodology, Formal analysis, Visualization, Writing --- Original Draft, \\
	Luis E.C.\ Rocha:\@ Conceptualization, Methodology, Writing --- Review \& Editing, Funding Acquisition, \\
	Tijs W.\ Alleman:\@ Investigation, Supervision, Validation, Writing --- Review \& Editing, \\
	Jan M.\ Baetens:\@ Conceptualization, Methodology, Writing --- Review \& Editing, Project Administration, Funding Acquisition,

	\newpage
	\section*{Glossary}
	\begin{tabular}{lcp{.645\textwidth}}
	\hline
	\textbf{Term} & \textbf{Symbol} & \textbf{Description} \\\hline
	Reddit       		  &         		& Social media platform where users can post links, images, and text. \\
	Subreddit    		  &         		& Community within Reddit, denoted by prefix r/. \\
	Submission   		  &         		& Top-level post on Reddit. \\
	Comment      		  &         		& Reply to a submission or other comment. \\
	Post         		  &         		& Submission or comment. \\
	Thread       		  &         		& A submission with all its comments. \\
	Parent       		  &         		& Submission or comment a comment is a reply to. \\
	Ancestors    		  &         		& All successive parents of a comment up to and including the submission. \\
	Discussion   		  &         		& Part of a thread with a common topic. \\
	Initiator    		  &         		& Creator of a post that does not share its topic with any of its ancestors. \\
	Participant  		  &         		& User in a discussion that is not the initiator. \\
	Ancestral context     & \( A^n_k \) & The \( n \) closest ancestors of a comment \( k \). \\
	User context 		  & \( U^n_k \) & The parents of comment \( k \) and of the \( n-1 \) preceding comments made by same author on the same topic. \\
	mbert-ctbt   		  &         		& mbert-corona-tweets-belgium-topics, the used topic model~\cite{Scott2021}. \\
	roberta-tbsl 		  &         		& twitter-roberta-base-sentiment-latest, the used sentiment model~\cite{Loureiro2022}. \\
	Sentiment    		  & \( s \) 		& The emotional tone of a text classified by roberta-tbsl. \\
	Homophily    		  & \( h \) 		& Tendency of individuals to associate with similar ones, here specifically with similar sentiment and calculated by Equation~\eqref{eq:homophily}. \\
	Consensus     		  &         		& A pattern of homophily where users predominantly express similar sentiment, leading to a unimodal distribution. \\
	Polarization  		  &         		& A pattern of homophily where two clusters of users express different sentiment, leading to a bimodal distribution. \\
	SLEBC        		  &         		& Smooth Latent-Expressed Bounded Confidence (Equation~\eqref{eq:model}). \\\hline
	\end{tabular}

	\section*{Declaration of generative AI and AI-assisted technologies in the writing process.}
	During the preparation of this text the authors used ChatGPT and Grammarly in order to improve the grammar, structure and flow of the text.
	After using these services, the authors reviewed and edited the content as needed and take full responsibility for the content of the published article.

	\newpage
	\bibliographystyle{elsarticle-num}
	\bibliography{arxiv.bib}

\begin{thebibliography}{10}
\expandafter\ifx\csname url\endcsname\relax
  \def\url#1{\texttt{#1}}\fi
\expandafter\ifx\csname urlprefix\endcsname\relax\def\urlprefix{URL }\fi
\expandafter\ifx\csname href\endcsname\relax
  \def\href#1#2{#2} \def\path#1{#1}\fi

\bibitem{Piette2022}
C.~Piette, J.~Tielens, How {B}elgian firms fared in the {COVID}-19 pandemic?, Nbb Economic Review (2022) 1--33\href {http://arxiv.org/abs/https://savearchive.zbw.eu/bitstream/11159/631033/1/1860602576_0.pdf} {\path{arXiv:https://savearchive.zbw.eu/bitstream/11159/631033/1/1860602576_0.pdf}}.

\bibitem{Reiriz2023}
M.~Reiriz, M.~Donoso-González, B.~Rodríguez-Expósito, S.~Uceda, A.~I. Beltrán-Velasco, Impact of {COVID}-19 confinement on mental health in youth and vulnerable populations: {A}n extensive narrative review, Sustainability 15~(4) (2023).
\newblock \href {https://doi.org/10.3390/su15043087} {\path{doi:10.3390/su15043087}}.

\bibitem{Alleman2026}
T.~W. Alleman, K.~Schoors, J.~M. Baetens, A dynamic disequilibrium input–output model of the {Belgian} {COVID}-19 pandemic, Economic Systems Research (2026) 1--25\href {https://doi.org/10.1080/09535314.2026.2622365} {\path{doi:10.1080/09535314.2026.2622365}}.

\bibitem{BroekAltenburg2021}
E.~van~den Broek-Altenburg, A.~Atherly, Adherence to {COVID}-19 policy measures: {Behavioral} insights from {The} {Netherlands} and {Belgium}, {PLOS} One 16~(5) (2021) e0250302.
\newblock \href {https://doi.org/10.1371/journal.pone.0250302} {\path{doi:10.1371/journal.pone.0250302}}.

\bibitem{Luyten2022}
J.~Luyten, E.~Schokkaert, Belgium’s response to the {COVID}-19 pandemic, Health Economics, Policy and Law 17~(1) (2022) 37--47.
\newblock \href {https://doi.org/10.1017/S1744133121000232} {\path{doi:10.1017/S1744133121000232}}.

\bibitem{West2020}
R.~West, S.~Michie, G.~J. Rubin, R.~Amlôt, Applying principles of behaviour change to reduce {SARS}-{CoV}-2 transmission, Nature Human Behaviour 4~(5) (2020) 451--459.
\newblock \href {https://doi.org/10.1038/s41562-020-0887-9} {\path{doi:10.1038/s41562-020-0887-9}}.

\bibitem{Krawczyk2021}
K.~Krawczyk, T.~Chelkowski, D.~J. Laydon, S.~Mishra, D.~Xifara, B.~Gibert, S.~Flaxman, T.~Mellan, V.~Schwämmle, R.~Röttger, J.~T. Hadsund, S.~Bhatt, Quantifying online news media coverage of the {COVID}-19 pandemic: {Text} mining study and resource, Journal of Medical Internet Research 23~(6) (2021) e28253.
\newblock \href {https://doi.org/10.2196/28253} {\path{doi:10.2196/28253}}.

\bibitem{Cinelli2020}
M.~Cinelli, W.~Quattrociocchi, A.~Galeazzi, C.~M. Valensise, E.~Brugnoli, A.~L. Schmidt, P.~Zola, F.~Zollo, A.~Scala, The {COVID}-19 social media infodemic, Scientific Reports 10~(1) (2020) 16598.
\newblock \href {https://doi.org/10.1038/s41598-020-73510-5} {\path{doi:10.1038/s41598-020-73510-5}}.

\bibitem{Cinelli2025}
M.~Cinelli, F.~Gesualdo, Infodemic versus viral information spread: {Key} differences and open challenges, JMIR Infodemiology 5~(1) (2025) e57455.
\newblock \href {https://doi.org/10.2196/57455} {\path{doi:10.2196/57455}}.

\bibitem{Alamoodi2021}
A.~H. Alamoodi, B.~B. Zaidan, A.~A. Zaidan, O.~S. Albahri, K.~I. Mohammed, R.~Q. Malik, E.~M. Almahdi, M.~A. Chyad, Z.~Tareq, A.~S. Albahri, H.~Hameed, M.~Alaa, Sentiment analysis and its applications in fighting {COVID}-19 and infectious diseases: {A} systematic review, Expert Systems with Applications 167~(114155) (2021).
\newblock \href {https://doi.org/10.1016/j.eswa.2020.114155} {\path{doi:10.1016/j.eswa.2020.114155}}.

\bibitem{Kurten2021}
S.~Kurten, K.~Beullens, \#{Coronavirus}: {Monitoring} the {Belgian} {Twitter} discourse on the severe acute respiratory syndrome coronavirus 2 pandemic, Cyberpsychology, Behavior, and Social Networking 24~(2) (2021) 117--122.
\newblock \href {https://doi.org/10.1089/cyber.2020.0341} {\path{doi:10.1089/cyber.2020.0341}}.

\bibitem{Park2024}
B.~Park, I.~S. Jang, D.~Kwak, Sentiment analysis of the {COVID}-19 vaccine perception, Health Informatics Journal 30~(1) (2024) 14604582241236131.
\newblock \href {https://doi.org/10.1177/14604582241236131} {\path{doi:10.1177/14604582241236131}}.

\bibitem{Lyu2022}
H.~Lyu, J.~Wang, W.~Wu, V.~Duong, X.~Zhang, T.~D. Dye, J.~Luo, Social media study of public opinions on potential {COVID}-19 vaccines: {I}nforming dissent, disparities, and dissemination, Intelligent Medicine 2~(1) (2022) 1--12.
\newblock \href {https://doi.org/10.1016/j.imed.2021.08.001} {\path{doi:10.1016/j.imed.2021.08.001}}.

\bibitem{Cheng2023}
T.~Cheng, B.~Han, Y.~Liu, Exploring public sentiment and vaccination uptake of {COVID}-19 vaccines in {England}: a spatiotemporal and sociodemographic analysis of {Twitter} data, Frontiers In Public Health 11 (2023) 1193750.
\newblock \href {https://doi.org/10.3389/fpubh.2023.1193750} {\path{doi:10.3389/fpubh.2023.1193750}}.

\bibitem{Pierri2023}
F.~Pierri, M.~R. DeVerna, K.-C. Yang, D.~Axelrod, J.~Bryden, F.~Menczer, One year of {COVID}-19 vaccine misinformation on {Twitter}: {Longitudinal} study, Journal of Medical Internet Research 25 (2023) e42227.
\newblock \href {https://doi.org/10.2196/42227} {\path{doi:10.2196/42227}}.

\bibitem{Cinelli2021}
M.~Cinelli, G.~De~Francisci~Morales, A.~Galeazzi, W.~Quattrociocchi, M.~Starnini, The echo chamber effect on social media, Proceedings of the National Academy of Sciences 118~(9) (2021) e2023301118.
\newblock \href {https://doi.org/10.1073/pnas.2023301118} {\path{doi:10.1073/pnas.2023301118}}.

\bibitem{Medford2020}
R.~J. Medford, S.~N. Saleh, A.~Sumarsono, T.~M. Perl, C.~U. Lehmann, An “{Infodemic}”: {L}everaging high-volume {Twitter} data to understand early public sentiment for the coronavirus disease 2019 outbreak, Open Forum Infectious Diseases 7~(7) (2020) ofaa258.
\newblock \href {https://doi.org/10.1093/ofid/ofaa258} {\path{doi:10.1093/ofid/ofaa258}}.

\bibitem{Lanier2022}
H.~D. Lanier, M.~I. Diaz, S.~N. Saleh, C.~U. Lehmann, R.~J. Medford, Analyzing {COVID}-19 disinformation on {Twitter} using the hashtags \#scamdemic and \#plandemic: {Retrospective} study, {PLOS} One 17~(6) (2022) e0268409.
\newblock \href {https://doi.org/10.1371/journal.pone.0268409} {\path{doi:10.1371/journal.pone.0268409}}.

\bibitem{Xie2023b}
T.~Y. Xie, Y.~R. Ge, Q.~Xu, S.~Chen, Public awareness and sentiment analysis of {COVID}-related discussions using {BERT}-based infoveillance, AI 4~(1) (2023) 333--347.
\newblock \href {https://doi.org/10.3390/ai4010016} {\path{doi:10.3390/ai4010016}}.

\bibitem{Scott2021}
K.~Scott, P.~Delobelle, B.~Berendt, Measuring shifts in attitudes towards {COVID}-19 measures in {Belgium}, Computational Linguistics in the Netherlands Journal 11 (2021) 161--171.

\bibitem{Medvedev2019}
A.~N. Medvedev, R.~Lambiotte, J.-C. Delvenne, The anatomy of {Reddit}: {A}n overview of academic research, in: F.~Ghanbarnejad, R.~Saha~Roy, F.~Karimi, J.-C. Delvenne, B.~Mitra (Eds.), Dynamics On and Of Complex Networks III, Springer International Publishing, 2019, pp. 183--204.
\newblock \href {https://doi.org/10.1007/978-3-030-14683-2_9} {\path{doi:10.1007/978-3-030-14683-2_9}}.

\bibitem{Corradini2024}
E.~Corradini, Deconstructing cultural appropriation in online communities: {A} multilayer network analysis approach, Information Processing \& Management 61~(3) (2024) 103662.
\newblock \href {https://doi.org/10.1016/j.ipm.2024.103662} {\path{doi:10.1016/j.ipm.2024.103662}}.

\bibitem{Melton2021}
C.~A. Melton, O.~A. Olusanya, N.~Ammar, A.~Shaban-Nejad, Public sentiment analysis and topic modeling regarding {COVID}-19 vaccines on the {Reddit} social media platform: {A} call to action for strengthening vaccine confidence, Journal of Infection And Public Health 14~(10) (2021) 1505--1512.
\newblock \href {https://doi.org/10.1016/j.jiph.2021.08.010} {\path{doi:10.1016/j.jiph.2021.08.010}}.

\bibitem{Liu2021}
Y.~Liu, C.~Whitfield, T.~Zhang, A.~Hauser, T.~Reynolds, M.~Anwar, Monitoring {COVID}-19 pandemic through the lens of social media using natural language processing and machine learning, Health Information Science And Systems 9~(1) (2021) 25.
\newblock \href {https://doi.org/10.1007/s13755-021-00158-4} {\path{doi:10.1007/s13755-021-00158-4}}.

\bibitem{Yurtsever2023}
M.~M.~E. Yurtsever, M.~Shiraz, E.~Ekinci, S.~Eken, Comparing {COVID}-19 vaccine passports attitudes across countries by analysing {Reddit} comments, Journal of Information Science (2023) 01655515221148356\href {https://doi.org/10.1177/01655515221148356} {\path{doi:10.1177/01655515221148356}}.

\bibitem{Yan2021}
C.~Yan, M.~Law, S.~Nguyen, J.~Cheung, J.~Kong, Comparing public sentiment toward {COVID}-19 vaccines across {C}anadian cities: {Analysis} of comments on {Reddit}, Journal of Medical Internet Research 23~(9) (2021) e32685.
\newblock \href {https://doi.org/10.2196/32685} {\path{doi:10.2196/32685}}.

\bibitem{Basile2021}
V.~Basile, F.~Cauteruccio, G.~Terracina, How dramatic events can affect emotionality in social posting: {T}he impact of {COVID}-19 on {R}eddit, Future Internet 13~(2) (2021) 29.
\newblock \href {https://doi.org/10.3390/fi13020029} {\path{doi:10.3390/fi13020029}}.

\bibitem{Hu2022}
M.~Hu, M.~Conway, Perspectives of the {COVID}-19 pandemic on {R}eddit: {C}omparative natural language processing study of the {U}nited {S}tates, the {U}nited {K}ingdom, {C}anada, and {A}ustralia, {JMIR} Infodemiology 2~(2) (2022) e36941.
\newblock \href {https://doi.org/10.2196/36941} {\path{doi:10.2196/36941}}.

\bibitem{Valensise2023}
C.~M. Valensise, M.~Cinelli, W.~Quattrociocchi, The drivers of online polarization: {Fitting} models to data, Information Sciences 642 (2023) 119152.
\newblock \href {https://doi.org/10.1016/j.ins.2023.119152} {\path{doi:10.1016/j.ins.2023.119152}}.

\bibitem{Kramer2014}
A.~D.~I. Kramer, J.~E. Guillory, J.~T. Hancock, Experimental evidence of massive-scale emotional contagion through social networks, Proceedings of the National Academy of Sciences 111~(24) (2014) 8788--8790.
\newblock \href {https://doi.org/10.1073/pnas.1320040111} {\path{doi:10.1073/pnas.1320040111}}.

\bibitem{DeFrancisciMorales2021}
G.~De~Francisci~Morales, C.~Monti, M.~Starnini, No echo in the chambers of political interactions on {Reddit}, Scientific Reports 11~(1) (2021) 2818.
\newblock \href {https://doi.org/10.1038/s41598-021-81531-x} {\path{doi:10.1038/s41598-021-81531-x}}.

\bibitem{Kozitsin2022}
I.~V. Kozitsin, A general framework to link theory and empirics in opinion formation models, Scientific Reports 12~(1) (2022) 5543.
\newblock \href {https://doi.org/10.1038/s41598-022-09468-3} {\path{doi:10.1038/s41598-022-09468-3}}.

\bibitem{Guilbeault2018}
D.~Guilbeault, J.~Becker, D.~Centola, Complex contagions: A decade in review, in: S.~Lehmann, Y.-Y. Ahn (Eds.), Complex Spreading Phenomena in Social Systems: Influence and Contagion in Real-World Social Networks, Springer International Publishing, 2018, pp. 3--25.
\newblock \href {https://doi.org/10.1007/978-3-319-77332-2_1} {\path{doi:10.1007/978-3-319-77332-2_1}}.

\bibitem{State2015}
B.~State, L.~Adamic, The diffusion of support in an online social movement: {Evidence} from the adoption of equal-sign profile pictures, in: Proceedings of the 18th {ACM} {Conference} on {Computer} {Supported} {Cooperative} {Work} \& {Social} {Computing}, {CSCW} '15, Association for Computing Machinery, 2015, pp. 1741--1750.
\newblock \href {https://doi.org/10.1145/2675133.2675290} {\path{doi:10.1145/2675133.2675290}}.

\bibitem{Baumann2020}
F.~Baumann, P.~Lorenz-Spreen, I.~M. Sokolov, M.~Starnini, Modeling echo chambers and polarization dynamics in social networks, Physical Review Letters 124~(4) (2020) 048301.
\newblock \href {https://doi.org/10.1103/PhysRevLett.124.048301} {\path{doi:10.1103/PhysRevLett.124.048301}}.

\bibitem{Haensch2023}
A.~Haensch, N.~Dragovic, C.~Borgers, B.~Boghosian, A geospatial bounded confidence model including mega-influencers with an application to {Covid}-19 vaccine hesitancy, Journal of Artificial Societies and Social Simulation 26~(1) (2023) 8.
\newblock \href {http://arxiv.org/abs/https://www.jasss.org/admin/get_pdf.php?source=https://www.jasss.org/26/1/8.html} {\path{arXiv:https://www.jasss.org/admin/get_pdf.php?source=https://www.jasss.org/26/1/8.html}}, \href {https://doi.org/10.18564/jasss.5027} {\path{doi:10.18564/jasss.5027}}.

\bibitem{Zeng2024}
R.~Zeng, X.~Chang, B.~Liu, Evolutionary modeling and analysis of opinion exchange and epidemic spread among individuals, Frontiers in Physics Volume 12 - 2024 (2024).
\newblock \href {https://doi.org/10.3389/fphy.2024.1501807} {\path{doi:10.3389/fphy.2024.1501807}}.

\bibitem{Baumgartner2020}
J.~Baumgartner, S.~Zannettou, B.~Keegan, M.~Squire, J.~Blackburn, The {Pushshift} {Reddit} dataset, Proceedings of the International Aaai Conference on Web and Social Media 14 (2020) 830--839.
\newblock \href {https://doi.org/10.1609/icwsm.v14i1.7347} {\path{doi:10.1609/icwsm.v14i1.7347}}.

\bibitem{Clauset2009}
A.~Clauset, C.~R. Shalizi, M.~E.~J. Newman, Power-law distributions in empirical data, SIAM Review 51~(4) (2009) 661--703.
\newblock \href {http://arxiv.org/abs/0706.1062} {\path{arXiv:0706.1062}}, \href {https://doi.org/10.1137/070710111} {\path{doi:10.1137/070710111}}.

\bibitem{Thukral2018}
S.~Thukral, H.~Meisheri, T.~Kataria, A.~Agarwal, I.~Verma, A.~Chatterjee, L.~Dey, Analyzing behavioral trends in community driven discussion platforms like {R}eddit, in: U.~Brandes, C.~Reddy, A.~Tagarelli (Eds.), 2018 IEEE/ACM International Confernce on Advances in Social Nnetworks Analysis and Mining (ASONAM), IEEE, 2018, pp. 662--669.

\bibitem{Papariello2024}
L.~Papariello, {xlm-roberta-base-language-detection}, Hugging Face, 10.57967/hf/2064 (2024).
\newblock \href {https://doi.org/10.57967/hf/2064} {\path{doi:10.57967/hf/2064}}.

\bibitem{Loureiro2022}
D.~Loureiro, F.~Barbieri, L.~Neves, L.~Espinosa~Anke, J.~Camacho-collados, {T}ime{LM}s: {D}iachronic language models from {T}witter, in: Proceedings of the 60th Annual Meeting of the Association for Computational Linguistics: System Demonstrations, Association for Computational Linguistics, 2022, pp. 251--260.
\newblock \href {https://doi.org/10.18653/v1/2022.acl-demo.25} {\path{doi:10.18653/v1/2022.acl-demo.25}}.

\bibitem{Bonifazi2022b}
G.~Bonifazi, F.~Cauteruccio, E.~Corradini, M.~Marchetti, L.~Sciarretta, D.~Ursino, L.~Virgili, A space-time framework for sentiment scope analysis in social media, Big Data And Cognitive Computing 6~(4) (2022) 130.
\newblock \href {https://doi.org/10.3390/bdcc6040130} {\path{doi:10.3390/bdcc6040130}}.

\bibitem{Huang2025}
X.~Huang, X.~Tang, Analyzing replies and interactions among users with different stances: {A} case study of the {Russia}-{Ukraine} conflict, in: X.~Tang, V.~N. Huynh, H.~Xia, Q.~Bai (Eds.), Knowledge and {Systems} {Sciences}, Springer Nature, 2025, pp. 109--123.
\newblock \href {https://doi.org/10.1007/978-981-96-0178-3_8} {\path{doi:10.1007/978-981-96-0178-3_8}}.

\bibitem{Alleman2021}
T.~W. Alleman, J.~Vergeynst, L.~De~Visscher, M.~Rollier, E.~Torfs, I.~Nopens, J.~M. Baetens, Assessing the effects of non-pharmaceutical interventions on sars-cov-2 transmission in belgium by means of an extended seiqrd model and public mobility data, Epidemics 37 (2021) 100505.
\newblock \href {https://doi.org/10.1016/j.epidem.2021.100505} {\path{doi:10.1016/j.epidem.2021.100505}}.

\bibitem{Sciensano2024}
Sciensano, {COVID-19}, \url{https://epistat.sciensano.be/covid/} (2024).

\bibitem{Bernal2017}
J.~L. Bernal, S.~Cummins, A.~Gasparrini, Interrupted time series regression for the evaluation of public health interventions: {A} tutorial, International Journal of Epidemiology 46~(1) (2017) 348--355.
\newblock \href {https://doi.org/10.1093/ije/dyw098} {\path{doi:10.1093/ije/dyw098}}.

\bibitem{Hegselmann2002}
R.~Hegselmann, U.~Krause, et~al., Opinion dynamics and bounded confidence models, analysis, and simulation, Journal of artificial societies and social simulation 5~(3) (2002).

\bibitem{Brooks2024a}
H.~Z. Brooks, P.~S. Chodrow, M.~A. Porter, Emergence of polarization in a sigmoidal bounded-confidence model of opinion dynamics, SIAM Journal on Applied Dynamical Systems 23~(2) (2024) 1442--1470.
\newblock \href {https://doi.org/10.1137/22M1527258} {\path{doi:10.1137/22M1527258}}.

\bibitem{DeGroot1974}
M.~H. DeGroot, Reaching a consensus, Journal of the American Statistical Association 69~(345) (1974) 118--121.
\newblock \href {https://doi.org/10.2307/2285509} {\path{doi:10.2307/2285509}}.

\bibitem{Ge2018}
H.~Ge, K.~Xu, Z.~Ghahramani, Turing: {A} language for flexible probabilistic inference, in: International Conference on Artificial Intelligence and Statistics, {AISTATS} 2018, 9-11 April 2018, Playa Blanca, Lanzarote, Canary Islands, Spain, 2018, pp. 1682--1690.

\bibitem{Hoffman2014}
M.~D. Hoffman, A.~Gelman, The {No}-{U}-{Turn} {Sampler}: {Adaptively} setting path lengths in {Hamiltonian} {Monte} {Carlo}, Journal of Machine Learning Research 15~(47) (2014) 1593--1623.

\bibitem{Watanabe2010}
S.~Watanabe, M.~Opper, Asymptotic equivalence of {B}ayes cross validation and widely applicable information criterion in singular learning theory, Journal of Machine Learning Research 11~(116) (2010) 3571--3594.
\newblock \href {http://arxiv.org/abs/https://www.jmlr.org/papers/volume11/watanabe10a/watanabe10a.pdf} {\path{arXiv:https://www.jmlr.org/papers/volume11/watanabe10a/watanabe10a.pdf}}.

\bibitem{Mann1947}
H.~B. Mann, D.~R. Whitney, On a test of whether one of two random variables is stochastically larger than the other, The Annals of Mathematical Statistics 18~(1) (1947) 50--60.
\newblock \href {https://doi.org/10.1214/aoms/1177730491} {\path{doi:10.1214/aoms/1177730491}}.

\bibitem{Hamed1998}
K.~H. Hamed, A.~Ramachandra~Rao, A modified {Mann}-{Kendall} trend test for autocorrelated data, Journal of Hydrology 204~(1) (1998) 182--196.
\newblock \href {https://doi.org/10.1016/S0022-1694(97)00125-X} {\path{doi:10.1016/S0022-1694(97)00125-X}}.

\bibitem{FOVVVL2004}
{Federale Overheidsdienst Volksgezondheid, Veiligheid van de Voedselketen en Leefmilieu}, Wet inzake experimenten op de menselijke persoon, eJustice, \url{https://www.ejustice.just.fgov.be/cgi/article_body.pl?language=nl&caller=summary&pub_date=04-05-18&numac=2004022376} (May 2004).

\bibitem{RedditSupport2025}
{Reddit Support}, Public {C}ontent {P}olicy, https://support.reddithelp.com/hc/en-us/articles/26410290525844-Public-Content-Policy (May 2025).

\bibitem{franzke2020}
a.~s. franzke, A.~Bechmann, M.~Zimmer, C.~Ess, the Association~of Internet~Researchers, Internet research: ethical guidelines 3.0, {AoIR} (2020).

\bibitem{Mancini2022}
A.~Mancini, A.~Desiderio, R.~Di~Clemente, G.~Cimini, Self-induced consensus of {Reddit} users to characterise the {GameStop} short squeeze, Scientific Reports 12~(1) (2022) 13780.
\newblock \href {https://doi.org/10.1038/s41598-022-17925-2} {\path{doi:10.1038/s41598-022-17925-2}}.

\bibitem{BinnenlandseZaken2020}
{Binnenlandse Zaken}, Ministerieel besluit houdende dringende maatregelen om de verspreiding van het coronavirus {COVID-19} te beperken, eJustice, \url{https://www.ejustice.just.fgov.be/eli/besluit/2020/03/13/2020030303/justel} (Mar. 2020).

\bibitem{Hale2021}
T.~Hale, N.~Angrist, R.~Goldszmidt, B.~Kira, A.~Petherick, T.~Phillips, S.~Webster, E.~Cameron-Blake, L.~Hallas, S.~Majumdar, H.~Tatlow, A global panel database of pandemic policies ({Oxford} {COVID}-19 {Government} {Response} {Tracker}), Nature Human Behaviour 5~(4) (2021) 529--538.
\newblock \href {https://doi.org/10.1038/s41562-021-01079-8} {\path{doi:10.1038/s41562-021-01079-8}}.

\bibitem{Wilmes2020}
S.~Wilmès, Start van fase 3 van het afbouwplan vanaf 8 juni, \href{https://web.archive.org/web/20200606095459/https://www.info-coronavirus.be/nl/news/nvr-0306/}{https://www.info-coronavirus.be/nl/news/nvr-0306/} (Jun. 2020).

\bibitem{Santens2020}
T.~Santens, Boete tot 1.600 euro voor wie avondklok schendt, sporten mag zonder masker: dit is nu vastgelegd in {Antwerpen}, VRT Nieuws\url{https://www.vrt.be/vrtnws/nl/2020/07/29/antwerpse-maatregelen/} (Jul. 2020).

\bibitem{Arnoudt2020}
R.~Arnoudt, Antwerpse maatregelen bijgestuurd: {H}oreca langer open, avondklok wordt nachtklok, evenementen weer mogelijk, VRT Nieuws, \url{https://www.vrt.be/vrtnws/nl/2020/08/12/antwerpse-gouverneur-maakt-een-nachtklok-van-de-avondklok/} (Aug. 2020).

\bibitem{DeCroo2020}
A.~De~Croo, {COVID}-19 alarmniveau gaat in: strengere regels vanaf maandag 19 oktober, \url{https://www.premier.be/nl/covid-19-alarmniveau-gaat-strengere-regels-vanaf-maandag-19-oktober} (Oct. 2020).

\bibitem{Overlegcomite2021}
Overlegcomité, Zomerplan: in vier stappen naar normaler levens, \url{https://www.belgium.be/nl/nieuws/2021/zomerplan_vier_stappen_naar_normaler_levens} (May 2021).

\bibitem{BinnenlandseZaken2021}
{Binnenlandse Zaken}, Koninklijk besluit houdende wijziging van het koninklijk besluit van 28 oktober 2021 houdende de nodige maatregelen van bestuurlijke politie teneinde de gevolgen voor de volksgezondheid van de afgekondigde epidemische noodsituatie betreffende de coronavirus {COVID-19} pandemie te voorkomen of te beperken., eJustice, \url{https://www.ejustice.just.fgov.be/eli/besluit/2021/11/27/2021043241/justel} (Nov. 2021).

\bibitem{Overlegcomite2022}
Overlegcomité, Code oranje vanaf 18 februari 2022: geen sluitingsuur horeca meer, nachtleven open, \url{https://www.belgium.be/nl/nieuws/2022/code_oranje_vanaf_18_februari_2022_geen_sluitingsuur_horeca_meer_nachtleven_open} (Feb. 2022).

\bibitem{TT2020}
TT, Mondmaskers vanaf zaterdag verplicht in alle winkels, bioscopen, gebedshuizen, bibliotheken en musea, HLN, \url{https://www.hln.be/binnenland/mondmaskers-vanaf-zaterdag-verplicht-in-alle-winkels-bioscopen-gebedshuizen-bibliotheken-en-musea~a8e6f1e8/} (Jul. 2020).

\bibitem{Overlegcomite2021a}
Overlegcomité, Overlegcomité bepaalt federale sokkel mondmaskerplicht, \url{https://www.belgium.be/nl/nieuws/2021/overlegcomite_bepaalt_federale_sokkel_mondmaskerplicht} (Sep. 2021).

\bibitem{Overlegcomite2021b}
Overlegcomité, Overlegcomité: voortaan brede mondmaskerplicht en verplicht telewerk, \url{https://www.belgium.be/nl/nieuws/2021/overlegcomite_voortaan_brede_mondmaskerplicht_en_verplicht_telewerk} (Nov. 2021).

\bibitem{BinnenlandseZaken2022}
{Binnenlandse Zaken}, Koninklijk besluit houdende wijziging van het koninklijk besluit van 28 oktober 2021 houdende de nodige maatregelen van bestuurlijke politie teneinde de gevolgen voor de volksgezondheid van de afgekondigde epidemische noodsituatie betreffende de coronavirus covid-19 pandemie te voorkomen of te beperken, eJustice, \url{https://www.ejustice.just.fgov.be/eli/besluit/2022/03/05/2022040532/justel} (Mar. 2022).

\bibitem{ThanhLe2020}
T.~Thanh~Le, Z.~Andreadakis, A.~Kumar, R.~Gómez~Román, S.~Tollefsen, M.~Saville, S.~Mayhew, The {COVID}-19 vaccine development landscape, Nature Reviews Drug Discovery 19~(5) (2020) 305--306.
\newblock \href {https://doi.org/10.1038/d41573-020-00073-5} {\path{doi:10.1038/d41573-020-00073-5}}.

\bibitem{Maerevoet2020}
E.~Maerevoet, Jos {Hermans} (96) krijgt als eerste {Vlaming} vaccin: "{Blij} dat ik spuit krijg, ik wil 100 worden", VRT Nieuws, \url{https://www.vrt.be/vrtnws/nl/2020/08/12/antwerpse-gouverneur-maakt-een-nachtklok-van-de-avondklok/} (Dec. 2020).

\bibitem{Arnoudt2021}
R.~Arnoudt, Derde prik voor bewoners woonzorgcentra, nog geen beslissing over thuiswonende 85-plussers, VRT Nieuws, \url{https://www.vrt.be/vrtnws/nl/2021/09/22/bewoners-van-woonzorgcentra-krijgen-derde-prik-coronavaccin/} (Sep. 2021).

\bibitem{Paelinck2021}
G.~Paelinck, Federale regering bereikt compromis over verplichte vaccinatie in de zorg, VRT Nieuws, \url{https://www.vrt.be/vrtnws/nl/2021/11/19/akkoord-vaccin/} (Nov. 2021).

\bibitem{Galindo2020}
G.~Galindo, 30\% of {Belgians} not in favour of getting coronavirus vaccine, \url{https://www.brusselstimes.com/129760/30-of-belgians-not-in-favour-of-getting-coronavirus-vaccine-survey} (Sep. 2020).

\bibitem{Kobayashi2022a}
R.~Kobayashi, Y.~Takedomi, Y.~Nakayama, T.~Suda, T.~Uno, T.~Hashimoto, M.~Toyoda, N.~Yoshinaga, M.~Kitsuregawa, L.~E.~C. Rocha, Evolution of public opinion on {COVID-19} vaccination in {J}apan: {L}arge-scale {T}witter data analysis, Journal of Medical Internet Research 24~(12) (2022) e41928.
\newblock \href {https://doi.org/10.2196/41928} {\path{doi:10.2196/41928}}.

\bibitem{Gozzi2020}
N.~Gozzi, M.~Tizzani, M.~Starnini, F.~Ciulla, D.~Paolotti, A.~Panisson, N.~Perra, Collective response to media coverage of the {COVID}-19 pandemic on {Reddit} and {Wikipedia}: {Mixed}-methods analysis, Journal of Medical Internet Research 22~(10) (2020) e21597.
\newblock \href {https://doi.org/10.2196/21597} {\path{doi:10.2196/21597}}.

\bibitem{Ancona2022}
C.~Ancona, F.~L. Iudice, F.~Garofalo, P.~De~Lellis, A model-based opinion dynamics approach to tackle vaccine hesitancy, Scientific Reports 12~(1) (2022) 11835.
\newblock \href {https://doi.org/10.1038/s41598-022-15082-0} {\path{doi:10.1038/s41598-022-15082-0}}.

\bibitem{Bedson2021}
J.~Bedson, L.~A. Skrip, D.~Pedi, S.~Abramowitz, S.~Carter, M.~F. Jalloh, S.~Funk, N.~Gobat, T.~Giles-Vernick, G.~Chowell, J.~R. de~Almeida, R.~Elessawi, S.~V. Scarpino, R.~A. Hammond, S.~Briand, J.~M. Epstein, L.~Hébert-Dufresne, B.~M. Althouse, A review and agenda for integrated disease models including social and behavioural factors, Nature Human Behaviour 5~(7) (2021) 834--846.
\newblock \href {https://doi.org/10.1038/s41562-021-01136-2} {\path{doi:10.1038/s41562-021-01136-2}}.

\bibitem{Alipour2024}
S.~Alipour, A.~Galeazzi, E.~Sangiorgio, M.~Avalle, L.~Bojic, M.~Cinelli, W.~Quattrociocchi, Cross-platform social dynamics: {A}n analysis of {ChatGPT} and {COVID}-19 vaccine conversations, Scientific Reports 14~(1) (2024) 2789.
\newblock \href {https://doi.org/10.1038/s41598-024-53124-x} {\path{doi:10.1038/s41598-024-53124-x}}.

\end{thebibliography}

	\newpage
	\begin{appendix}
		\section{Supporting Equations}\label{app:equations}

		\subsection*{Interrupted Time Series Analysis (Section~\ref{sec:met-temporal})}

		The linear interrupted time series analysis uses an augmented design matrix to find changes in level (\( \beta_0 \)) and slope (\( \beta_1 \)) after each designated dates.
		Let \( t \) denote the time (in days) since the start of the time series, and let \( \{d_1, d_2, \ldots, d_m\} \) be the set of event dates.
		The design matrix \( \mathbf{X} \) is constructed as:
		\begin{equation}
			\mathbf{X} = \left[ \mathbf{1} \quad \mathbf{t} \quad \mathbf{D}_1 \quad (\mathbf{D}_1 \circ (\mathbf{t} - d_1)) \quad \cdots \quad \mathbf{D}_m \quad (\mathbf{D}_m \circ (\mathbf{t} - d_m)) \right],
		\end{equation}
		where \( \mathbf{1} \) is a vector of ones, \( \mathbf{t} \) is the time vector, \( \mathbf{D}_j \) is the indicator vector with \( [\mathbf{D}_j]_t = 1 \) if \( t \geq d_j \) and 0 otherwise, \( \circ \) denotes element-wise multiplication, and columns are concatenated.
		Ordinary least squares regression of daily post counts \( \mathbf{y} \) on \( \mathbf{X} \) yields coefficients \( \boldsymbol{\beta} = (\mathbf{X}^T \mathbf{X})^{-1} \mathbf{X}^T \mathbf{y} \).
		For each event \( j \), the level change \( \Delta\beta_{0,j} \) is the coefficient of \( \mathbf{D}_j \) and the slope change \( \Delta\beta_{1,j} \) is the coefficient of the interaction term \( \mathbf{D}_j \circ (\mathbf{t} - d_j) \).
		Confidence intervals are computed from the covariance matrix of \( \boldsymbol{\beta} \).

		\subsection*{Weighted Sentiment (Section~\ref{sec:met-temporal})}

		For each topic \( x \), the weighted sentiment set is defined as
		\begin{equation}\label{eq:Sx}
			S^x = \left\{ v_k s_k \mid \text{post } k \text{ is about topic } x \right\},
		\end{equation}
		where \( v_k \) is the score and \( s_k \in [-1, 1] \) is the sentiment of post \( k \).
		The per-day subset is
		\begin{equation}\label{eq:Sxd}
			S^x_d = \left\{ \tilde{s}_k \in S^x \mid \text{post } k \text{ was created on date } d \right\}.
		\end{equation}

		\subsection*{Topic Contagion Null Model (Section~\ref{sec:met-activity})}

		Given \( n_I \) initiations and \( n_P \) participations in a user's posting sequence, the total number of orderings under the null model (uniform over all sequences) is
		\begin{equation}
			N(n_I, n_P) = \binom{n_I + n_P}{n_I}.
		\end{equation}
		The number of sequences in which the \( i \)-th post is the first initiation (requiring the first \( i-1 \) posts to be participations, the \( i \)-th to be an initiation, and the rest in any order) is
		\begin{equation}
			T(i|n_I, n_P) = \binom{n_I + n_P - i}{n_I - 1}.
		\end{equation}
		The null probability is therefore
		\begin{equation}\label{eq:ecdf}
			P(i|n_I, n_P) = \frac{T(i|n_I, n_P)}{N(n_I, n_P)}.
		\end{equation}

		\subsection*{Alternative Models (Section~\ref{sec:met-model})}

		The linear alternative model updates expressed sentiment \( \bar{e}_i \) as a weighted average of the latent sentiment \( u_i \) and the parent comment sentiment \( s_p \):
		\begin{equation}\label{eq:linear}
			L_{\alpha}(s_1, s_2) = s_1 + \alpha (s_2 - s_1).
		\end{equation}

		The stateless alternative model removes the latent sentiment state \( u_i \), updating expressed sentiment \( \bar{e}_i \) directly based on the set of received and replied-to sentiment:
		\begin{equation}\label{eq:stateless}
			\bar{e}_i[t] \sim \tN{\bigotimes_{e_{k}\in I_i[t-1, t]} B_{\alpha_i, \epsilon_i}(\bar{e}_i[t-1], e_{k})}{\sigma_i}.
		\end{equation}

		\subsection*{Watanabe-Akaike Information Criterion (Section~\ref{sec:met-model})}

		The SLEBC model fit was quantified using the Watanabe-Akaike Information Criterion~\cite{Watanabe2010}.
		Given \( M \) samples \( \theta_m \) of the posterior distribution containing model parameters \( \alpha_u, \alpha_e, \epsilon, \sigma \), and \( N \) observed sentiment \( s_j \), the log pointwise predictive density is
		\begin{equation}\label{eq:waic}
			\text{lppd} = \sum_{j=1}^N \log \left( \frac{1}{M} \sum_{m=1}^M p(s_j | \theta_m) \right).
		\end{equation}
		This log likelihood increases if the sampled parameters better predict the observed sentiment.
		A penalty term, the effective number of parameters \( n_{eff} \), is added to account for model complexity, calculated as the sum of the variances of the log likelihood across posterior samples,
		\begin{equation}
			n_{eff} = \sum_{j=1}^N \text{Var}_{m=1}^M \left( \log p(s_j | \theta_m) \right).
		\end{equation}
		The WAIC is then computed as
		\begin{equation}
			\text{WAIC} = -2 (\text{lppd} - n_{eff}).
		\end{equation}
		Lower WAIC values indicate better model fit, balancing predictive accuracy and complexity.
		Asymptotically, WAIC is equivalent to leave-one-out cross-validation loss.

	\section{Sensitivity Analysis of the Homophily Measure}\label{app:sensitivity}

	\begin{table}[bhp]
		\centering
		\begin{tabular}{lcccccc}
			\hline
			& \multicolumn{2}{c}{\emph{Lockdowns}} & \multicolumn{2}{c}{\emph{Masks}} & \multicolumn{2}{c}{\emph{Vaccination}} \\
			\( w_H \) & \( h|w_H \) & \( \delta h \) & \( h|w_H \) & \( \delta h \) & \( h|w_H \) & \( \delta h \) \\ \hline
			0.2     & 0.24468 & -0.00626 & 0.21223 & -0.00052 & 0.14548 & 0.01298 \\
			0.1     & 0.24571 & -0.00209 & 0.21232 & -0.0001  & 0.1446  & 0.00686 \\
			0.05    & 0.24614 & -0.00033 & 0.21205 & -0.00138 & 0.14387 & 0.00181 \\
			0.025   & 0.24586 & -0.00146 & 0.212   & -0.00157 & 0.14304 & -0.00401 \\
			0.00125 & 0.24622 & 0        & 0.21234 & 0        & 0.14361 & 0 \\\hline
		\end{tabular}

		\caption{%
			Sensitivity analysis of homophily measure \( h \) with respect to input bin width \( w_H \).
		}
		\label{tab:sensitivity}
	\end{table}

	As the homophily measure, \( h \) depends on the bin width \( w_H \), we conducted a sensitivity analysis by varying \( w_H \) and observing the effects on \( h \).
	We calculated \( h \) for different values of \( w_H \) as well as its relative difference with the smallest one,
	\begin{equation}
		\delta h(w_H) = \frac{h|_{w_H} - h|_{0.00125}}{h|_{0.00125}}.
	\end{equation}
	Table~\ref{tab:sensitivity} presents the results of this analysis.
	While the effect of varying \( w_H \) depends on the topic, we conclude that \( w_H=0.05 \) is an appropriate choice as it provides a good scale to interpret the figures and does not differ more than 0.2\% from the smallest tested bin width.
	\end{appendix}

\end{document}